\documentclass[aps,preprint,nofootinbib,tightenlines]{revtex4}

\usepackage{graphicx}  %
\usepackage{amsmath}
\usepackage{bm}  %

\begin{document}




\title{$\bm{J/\psi}$ and $\bm{\Upsilon}$ Polarization in Hadronic Production Processes\\
~}
\author{Eric Braaten}
\affiliation{Department of Physics, The Ohio State University,
Columbus, OH 43210, USA}
\author{James Russ}
\affiliation{Physics Department, Carnegie-Mellon University,
Pittsburgh, PA 15213, USA\\}

\keywords{Quarkonium, Polarization, Hadroproduction}

\begin{abstract}
Both charm and bottom quarks form nonrelativistic bound states
analogous to positronium.  The $J/\psi$ and $\psi(2S)$ charmonium states
and the first three $\Upsilon(nS)$ bottomonium states, all spin-triplet $S$-wave quarkonium states below open-heavy-flavor thresholds,
have relatively large branching ratios to $e^-e^+$ or $\mu^- \mu^+$ pairs.
In hadron collisions, experiments measuring lepton pairs can determine polarization by using angular correlation techniques.
The polarization, in turn, can be related theoretically to the production
mechanism for the bound state.
This review summarizes experimental studies with proton beams at fixed
target and colliding beam accelerators,
covering a center-of-mass energy range from 39 to 7000 GeV for
nucleon and antiproton targets.
Analyses using various polarization frames and spin-quantization axes
are described and results compared.
A pattern emerges that connects experimental results over the whole energy
span.
The theoretical implications of the pattern are presented and a set of new
measurements is proposed.\\

\noindent
Prepared for Volume 64 of the Annual Review of Nuclear and Particle Science
\end{abstract}

\maketitle

\section{INTRODUCTION}

Photon {\it polarization} has long been used to determine the properties 
of electromagnetic transition matrix elements.  
Measuring the polarization properties of the Cosmic Background Radiation 
is expected to provide strong constraints on models of the early universe~\cite{kosowsky}.  
Polarization measurements may play a crucial role in determining the spin 
of any new heavy particles discovered at the Large Hadron Collider.  
The power of polarization measurements is that they can follow changes in the matrix elements 
that contribute to a complex process as kinematic variables change. 

\begin{figure}
\centerline{ \includegraphics*[width=16cm,clip=true]{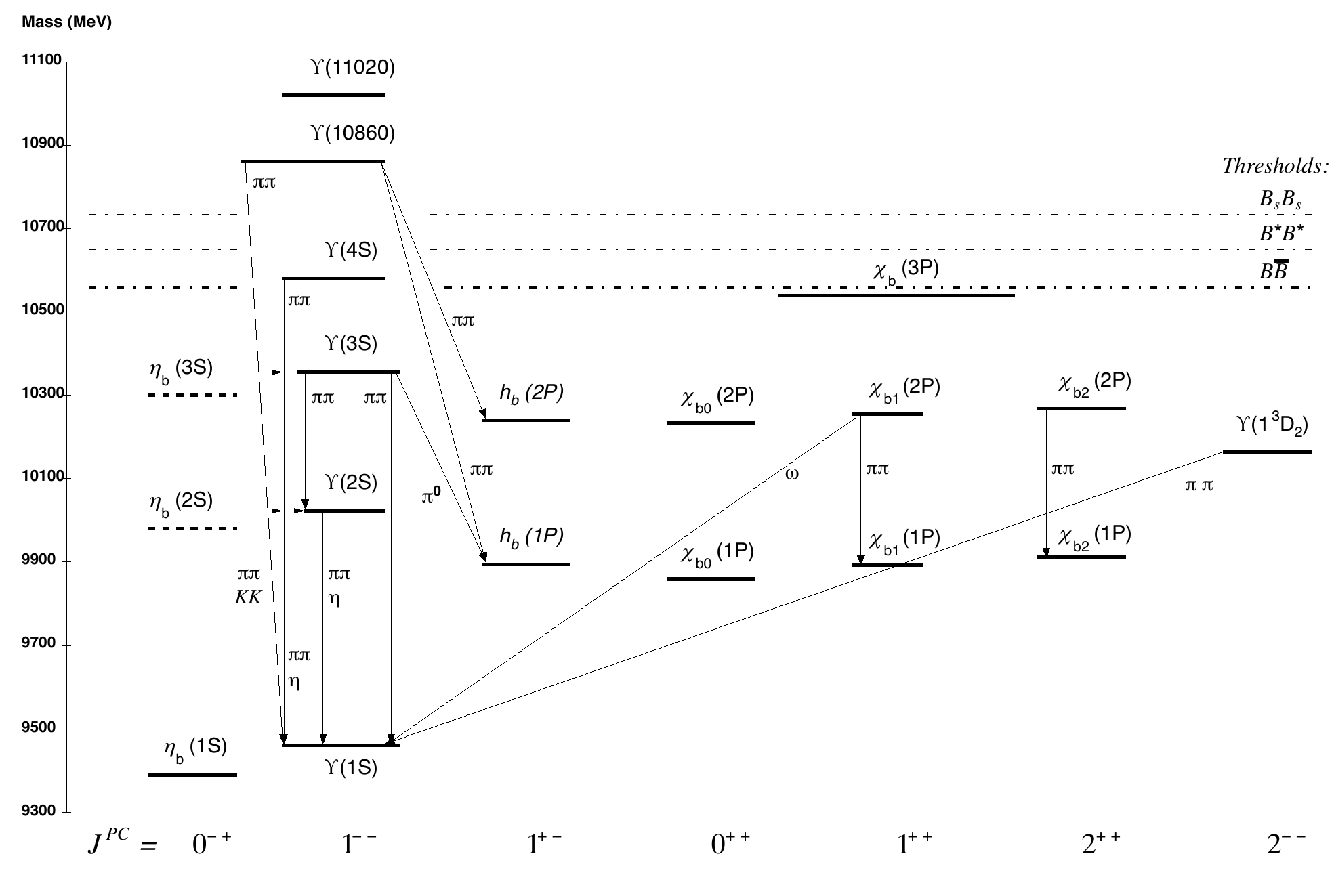}}
\caption{Bottomonium bound states and their observed hadronic transitions.  
\label{spectra}}
\end{figure}

This review treats recent developments in polarization measurements for {\it quarkonium} systems produced in hadronic collisions. These mesons are bound states of heavy-flavor quarks, $c\overline{c}$ or $b\overline{b}$.  They exhibit a positronium-like series of excited states with increasing principal and orbital-angular-momentum quantum numbers, 
shown for bottomonium in Fig.~\ref{spectra}. 
The spin-triplet $S$-wave states of quarkonium,
which have $J^{PC}$ quantum numbers $1^{--}$, can decay into a pair of leptons
through a virtual photon.  The dilepton angular distribution can be used 
as a probe of the quarkonium polarization in hadronic production processes.

The goal of  
quarkonium polarization measurements in high-energy hadron collisions is 
to determine the mechanisms by which a heavy quark-antiquark pair is produced 
by parton-parton collisions and by which it subsequently 
binds into a colorless meson $H$.  The relevant kinematic variables for inclusive production of $H$ 
are its {\it transverse momentum} $p_T$
and a longitudinal variable,\footnote{
Rapidity $y$ and Feynman $x_F$ are different longitudinal kinematic variables
for a hadron:
$y  \equiv \frac{1}{2} \ln[(E + \vec{p} \cdot \hat{z})/(E - 
\vec{p} \cdot \hat{z})]$, where $E$ and $\vec{p}$ are its energy 
and momentum in the center-of-momentum frame of the colliding
 hadrons and $\hat{z}$ is the collision axis,
while $x_F = \vec{p} \cdot \hat{z}/p_{\rm max}$.  At $y = 0$, $x_F = 0$.}
either its {\it rapidity} $y$ or its {\it Feynman variable} $x_F$.
Despite the complexity of describing strong production processes using 
{\it Quantum Chromodynamics} (QCD),  two simplifications 
arise for quarkonium production due to the heavy quark's mass
being much larger than the light-quark interaction energy scale 
$\Lambda_{\rm QCD}$.  
First, the {\it asymptotic freedom}  of the QCD coupling constant
allows the creation of the heavy quarks to be described perturbatively, 
provided also that the quarkonium state $H$ has sufficient transverse momentum 
to suppress interactions of the heavy quarks with the light-parton remnants of the colliding hadrons.
Second, the nonrelativistic velocities of the heavy quarks in the rest frame of $H$ 
simplify the nonperturbative dynamics of the formation of the bound state. 
The spin structure of the matrix element for the parity-conserving electromagnetic 
decay into leptons is completely determined by Lorentz invariance.
The decay angular distribution of the leptons carries information 
about the spin density matrix for the vector quarkonium 
and therefore about the process by which that quarkonium was produced.
Experiments can determine the 
variation of the decay angular distribution with kinematic variables. 
 Theory has to provide the interpretive framework to relate the data 
to the dominant matrix elements for the production process.  This 
review will consider the present state of both sides of this 
interpretive equation.

We recall that precision experiments are difficult. 
The record of experimental science shows that the first measurements of 
any important observable are sometimes not consistent with later, 
more sophisicated measurements made with larger data samples.  
We will review the available polarization experiments and their 
phase space coverage with an eye toward identifying the strengths 
of each.  At the end, we will summarize the situation and identify
what we know experimentally about quarkonium polarization.  
We give potential theoretical implications of the global trends of
the measurements to date.  Finally, we will review important next 
steps to be taken to improve our knowledge of the hadroproduction 
of quarkonium.

\section{POLARIZATION FRAMES}
\label{frames}

The polarization of a spin-triplet $S$-wave quarkonium state can be revealed 
by the angular distribution of the lepton pair into which it decays.
In this section, we discuss the polarization frames that can be used to 
define that angular distribution.

\subsection{Angular Distributions}
\label{sec:angular}

In the rest frame of a vector meson, its spin component along any 
{\it spin-quantization axis} has three possible eigenvalues: $m_s = 0, \pm 1$ 
(in units of $\hbar$).  The $m_s = 0$ state is called {\it longitudinal}
and the $m_s = \pm 1$ states are called {\it transverse}.
Because of the parity symmetry of QCD,
a collision of unpolarized hadrons cannot produce a vector meson with 
different probabilities for the spin states $+1$ and $-1$.  
We will consider a vector meson to have a net {\it polarization} if 
the probabilities for a single transverse spin state and the longitudinal 
spin state differ.\footnote{
Some physicists reserve the term {\it polarization} for
situations in which the spin states $+1$ and $-1$ have unequal probabilities.  
Situations in which the probabilities for $+1$ and $-1$ are equal
 but different from that for $m_s = 0$ are called {\it spin alignment}.}

When a vector meson decays into a lepton pair, its polarization
is reflected in the angular distribution of the leptons, as 
specified, e.g., in terms of the spherical angles
$\theta $ and $\phi$ for the momentum vector of the 
positively charged lepton in the rest frame of the vector meson.
In order to define those angles, it is necessary to choose a 
{\it polarization frame}.  The angle 
$\theta$ is the polar angle with respect to the spin-quantization 
axis.  An orthogonal axis in the collision plane must be specified
to define the zero of the azimuthal angle. 
For inclusive hadroproduction of a vector meson,
the only available vectors are the momenta of the vector meson and the colliding hadrons.
In the rest frame of the vector meson, the two orthogonal  axes lie in the collision
plane defined by the boosted momenta of the colliding hadrons.

A thorough discussion of the dilepton angular distribution 
and quarkonium polarization
has been presented by Faccioli {\it et al.}~\cite{fac1}.
The most general angular distribution for the dileptons from the 
decay of 
vector mesons produced by parity-invariant interactions is 
specified by three polarization parameters: 
$\lambda_\theta$, $\lambda_\phi$, and $\lambda_{\theta \phi}$.
The two-dimensional angular distribution is 
\begin{equation}
\frac{d W}{d (\cos \theta) d \phi} = \frac{3
[1 + \lambda_\theta \cos^2\theta + \lambda_\phi \sin^2\theta \cos(2 \phi)
 + \lambda_{\theta \phi} \sin(2\theta) \cos \phi ]}{4\pi(3+\lambda_\theta)} .
\label{dW-dtheta,dphi}
\end{equation}
The angular distribution has been normalized so that it integrates to 1.
The general constraints on the three polarization parameters are 
$|\lambda_\theta| \le 1$, $|\lambda_\phi| \le (1 + \lambda_\theta)/2$, and 
$\lambda_{\theta \phi}^2 \le  (1-\lambda_\theta)(1+\lambda_\theta - 2  \lambda_\phi)/4$ 
~\cite{Palestini:2010xu}. 

The three polarization variables $\lambda_\theta, \lambda_\phi,$ and 
$\lambda_{\theta \phi} $
can be determined from measurements of the one-dimensional distributions
obtained by projecting the two-dimensional distribution in 
Eq.~\ref{dW-dtheta,dphi} 
onto $\cos\theta$, $\phi$, or another angle defined by
$\tilde{\phi} = \phi - \mbox{$\frac14$} \pi [ 2 - {\rm sign}(\cos 
\theta)]$ ~\cite{fac1}:

\begin{eqnarray}
\frac{d W}{d (\cos \theta)} &=& \frac{3}{2(3+\lambda_\theta)}
\big[ 1 + \lambda_\theta \cos^2\theta \big] ,
\label{dW-dtheta}
\\
\frac{d W}{ d \phi} &=& \frac{1}{2\pi(3+\lambda_\theta)}
\big[ 3 + \lambda_\theta  + 2 \lambda_\phi \cos(2 \phi) \big],
\label{dW-dphi}
\\
\frac{d W}{d \tilde\phi} &=& \frac{1}{2\pi(3+\lambda_\theta)}
\big[ 3 + \lambda_\theta  + \sqrt2\, \lambda_{\theta \phi} 
\cos \tilde\phi\, \big] .
\label{dW-dphitilde}
\end{eqnarray}
In experiments with low statistics,  it may be necessary to
determine $\lambda_\theta$, $\lambda_\phi$, and $\lambda_{\theta \phi}$ 
from measurements of the three separate one-dimensional distributions
in Eqs.~\ref{dW-dtheta}, \ref{dW-dphi}, and \ref{dW-dphitilde}
in order to obtain stable results.
In experiments with higher statistics, the three polarization parameters 
can be determined with smaller systematic errors 
from measurements of the two-dimensional distribution in 
Eq.~\ref{dW-dtheta,dphi}.
The statistics in those measurements can be improved by exploiting two
 symmetries of the angular distribution in Eq.~\ref{dW-dtheta,dphi}: 
$\phi \rightarrow -\phi$ and
$(\theta, \phi) \rightarrow (\pi - \theta, \pi - \phi)$.   They 
allow the 
two-dimensional distribution to be folded into the first quadrant, 
{\it as long as the apparatus acceptance and efficiency 
are also symmetric under these two operations.}

A different choice for the polarization frame can be obtained by a rotation 
of the collision plane in the rest frame of the vector meson. 
The angular distribution in the new frame
has the same general form as in Eq.~\ref{dW-dtheta,dphi}, but with
different polarization parameters 
$\lambda_\theta$, $\lambda_\phi$, and $\lambda_{\theta \phi}$
that are functions of the old polarization parameters and the 
rotation angle.
There are combinations of the polarization
parameters that are independent of this rotation angle:
\begin{eqnarray}
\tilde \lambda &=& 
\frac{\lambda_\theta  + 3 \lambda_\phi}{1 - \lambda_\phi},
\label{lambda-tilde}
\\
{\tilde \lambda}' &=& 
\frac{(\lambda_\theta - \lambda_\phi)^2 + 4 \lambda_{\theta\phi}^2}
        {(3 + \lambda_\theta)^2}.
\label{newlambda-tilde}
\end{eqnarray}
The invariance of $\tilde \lambda$ was pointed out by 
Faccioli {\it et al.}~\cite{fac3,Faccioli:2010ji}.
The invariance of ${\tilde \lambda}'$, 
which depends on all three polarization parameters,
was pointed out by  Palestini \cite{Palestini:2010xu}.
These two frame-invariant polarization parameters provide 
powerful constraints on the accuracy of measurements 
of the dilepton angular distributions in 
different polarization frames.  
They could also be useful in theoretical calculations to check 
whether error estimates on predictions of the polarization are 
underestimated.

Because the same data are used in 
evaluating polarization variables in each frame, comparing the 
frame-independent quantities from two frames cannot be based on 
the statistical uncertainties, which are highly correlated.  
Generally, a Monte Carlo study of the range of expected variation 
is used to determine the consistency of comparisons of 
$\tilde{\lambda}$ or $\tilde{\lambda}'$ for different frames.

\subsection{Specific Polarization Frames}
\label{sec:choice}

The polarization frame can be specified by the direction of the 
spin-quantization axis in the plane containing the momentum vectors  
of the colliding hadrons in the quarkonium rest frame. 
There are several choices for the spin-quantization axis in the 
literature:
\begin{itemize}
\item
{\bf Gottfried-Jackson} (GJ) axis~\cite{GJ:1964}:
 the direction of the momentum  of one of the two colliding hadrons,
\item
{\bf Collins-Soper} (CS) axis~\cite{CS:1977}: the direction of the difference 
between the velocity vectors of the colliding hadrons,
\item
{\bf center-of-mass helicity} (cm-helicity)  axis:  
the direction of the boost required to go from the quarkonium rest frame 
to the center-of-momentum frame of the colliding hadrons,
\item
{\bf perpendicular helicity} ($\perp$-helicity) 
axis~\cite{Braaten:2008mz,Braaten:2008xg}:
the direction of the sum of the velocity vectors of the colliding hadrons
or, alternatively, the direction of the boost required to go from 
the quarkonium rest frame 
to the frame in which the quarkonium momentum is perpendicular to 
the axis of the colliding hadrons.
\end{itemize}

For some of these frames, there are simple physical mechanisms 
that tend to produce polarization 
in spin-triplet $S$-wave quarkonium~\cite{fac1,Braaten:2008xg}.
If the $Q \bar Q$ pair is created directly by a virtual gluon or 
virtual photon from the collision of a massless quark and antiquark 
that are collinear with the colliding hadrons,
the polarization will tend to be transverse in the CS frame
and longitudinal in the cm-helicity frame.
The transverse polarization in the CS frame 
follows from the helicity conservation of the interaction 
of the virtual gluon or photon with the light quark and antiquark.
If the $Q \bar Q$ pair is created directly by a virtual 
gluon or virtual photon with transverse momentum
that is much larger than the heavy quark mass,
the polarization of quarkonium will tend to be transverse 
in the cm-helicity and $\perp$-helicity frames.
The transverse polarization follows from the approximate helicity 
conservation of the interaction 
of the almost on-shell gluon or photon with the heavy quark and 
antiquark.

The polarization parameter $\lambda_\theta$ measures the degree 
of polarization with respect to the spin-quantization axis.
It can be expressed as 
$ (\sigma_T - 2 \sigma_L)/(\sigma_T + 2 \sigma_L)$,
where $\sigma_T$ and $\sigma_L$ are the cross sections 
for the two transverse states and for the single longitudinal state, 
respectively. 
A vector meson can be polarized with respect to one quantization axis
and unpolarized with respect to another.
Measurements of  $\lambda_\theta$  with respect to
two orthogonal spin-quantization axes carries much more information 
about the polarization mechanism
than a single measurement~\cite{fac2}.
If $\lambda_\theta = 0$ for both frames, then  $\lambda_\phi = 0$
for both frames and $\lambda_{\theta\phi}$ is equal and opposite in the two frames.
At zero rapidity, the CS axis is orthogonal to the cm-helicity axis,
which coincides with the $\perp$-helicity axis.
The CS and  $\perp$-helicity axes remain orthogonal at nonzero rapidity,
so measurements of $\lambda_\theta$ with respect to these two axes will
provide the most information about the polarization 
mechanism~\cite{Braaten:2008mz,Braaten:2008xg}.

\section{THEORETICAL CONSIDERATIONS}
\label{theo}

In this section, we describe various theoretical approaches to
quarkonium production in QCD and discuss their implications for polarization.

\subsection{General considerations}

Because of the asymptotic freedom of the QCD coupling constant,
amplitudes involving a large momentum transfer $Q$
can be calculated using perturbative QCD (pQCD) 
as an expansion in powers of $\alpha_s(Q)$, 
provided there is a factorization theorem that guarantees 
the insensitivity of the amplitude to much smaller momentum scales.
The creation of a $Q \bar Q$ pair in a collision of light partons
involves a momentum transfer of order $m_Q$,
where $m_Q$ is the heavy-quark mass.
The creation of a $Q \bar Q$ pair with transverse momentum $p_T$
much larger than $m_Q$ involves a momentum transfer of order $p_T$.
In an amplitude involving large $p_T$, some factors of the QCD 
coupling constant should be  $\alpha_s(p_T)$,
while others may more appropriately be $\alpha_s(m_Q)$.
If the momentum scales $m_Q$ and $p_T$ are not separated,
the momentum scales in all factors of $\alpha_s$ are usually set to a 
common value, such as $(m_Q^2+p_T^2)^{1/2}$.
In a pQCD calculation,
we will refer to the first few terms in the expansion in powers of $\alpha_s$
as leading order (LO), next-to-leading order (NLO), and 
next-to-next-to-leading order (N$^2$LO).

When $p_T > m_Q$, a pQCD cross section for producing a $Q \bar Q$ pair 
with small relative momentum can, up to logarithms of $p_T/m_Q$, 
be expanded in powers of $m_Q/p_T$, where $m_Q$ is the heavy-quark mass.
Because QCD has asymptotic scale invariance at large momentum transfer,
the leading power (LP) in $d \sigma/dp_T^2$ must, by dimensional analysis, be $1/p_T^4$.
In a cross section that is summed over the quarkonium spin states,
the next-to-leading power (NLP) is $m_Q^2/p_T^6$.
The cross sections for individual quarkonium spin states can also have 
the intermediate power $m_Q/p_T^5$.

The formation of a quarkonium $H$ from a $Q \bar Q$ pair
with small relative momentum is an inherently nonperturbative process,
but there are simplifications that arise from the $Q$ and $\bar Q$
being nonrelativistic in the rest frame of $H$.
The relative importance of nonperturbative transitions of the 
$Q \bar Q$ pair is determined by how their amplitudes scale with the 
relative velocity $v$ of the $Q$ and $\bar Q$.
The typical relative velocity of the $c \bar c$ pair in the $J/\psi$ is 
given roughly by $v^2 \approx 0.3$.
The typical relative velocity of the $b \bar b$ pair in the $\Upsilon(1S)$ is 
given roughly by $v^2 \approx 0.1$.
The typical relative velocities in the radially excited states are larger,
but they may still be small enough to allow scaling with $v$
to be useful as an organizing principle for nonperturbative transitions
of the $Q \bar Q$ pair.

Nonrelativistic QCD (NRQCD) is an effective field theory for the sector of QCD 
that includes a nonrelativistic heavy quark and antiquark. 
The Lagrangian for NRQCD includes infinitely many terms,
but they can be organized according to how their contributions to the energy
of quarkonium scale with the typical relative velocity 
of the $Q \bar Q$ pair  \cite{Lepage:1992tx}.  
The leading terms of order $v^2$ give splittings
between the radial and orbital-angular-momentum excitations of quarkonium. 
The terms of order $v^4$ give splittings within orbital-angular-momentum
multiplets.  By including terms of increasingly higher order in $v$,
the spectrum of quarkonium in QCD can be  reproduced with
increasingly higher accuracy.  NRQCD can also be used to organize 
nonperturbative effects in the annihilation decays of quarkonium 
and in the inclusive production of quarkonium \cite{Bodwin:1994jh}.

An important qualitative feature of the nonrelativistic dynamics of a heavy quark 
is the suppression of spin flip.  
The amplitudes for transitions of the $Q \bar Q$ pair  in which the spin state 
of the $Q$ or $\bar Q$ changes are suppressed by a factor of $v^2$. 
Because of this suppression, the hadronic and electromagnetic
transitions of an excited spin-triplet quarkonium state 
are primarily to lower spin-triplet states.  
The suppression of spin flip for a heavy quark also has implication for production of quarkonium.  
If the parton collisions that create a $Q \bar Q$ pair with small relative momentum
give it a nonzero polarization, the subsequent binding of 
the $Q \bar Q$ pair will tend to preserve its spin state,
passing the  polarization on to the quarkonium.

\subsection{Theoretical approaches}
\label{sec:general}

\subsubsection{NRQCD Factorization Formula}
\label{sec:nrqcd-fact}

The NRQCD factorization formula \cite{Bodwin:1994jh} 
is a conjectured factorization formula 
in which nonperturbative effects associated with the binding of a $Q \bar Q$ pair 
into quarkonium are organized into multiplicative constants. 
The theoretical status of the conjecture is discussed in Ref.~\cite{Bodwin:2013nua}.
The NRQCD factorization formula states that the inclusive cross section 
for producing a quarkonium state $H$ 
in the collision of the light hadrons $A$ and $B$ can be expressed as the sum of 
inclusive pQCD cross sections for producing a $Q\bar Q$ pair
multiplied by {\it NRQCD matrix elements}:
\begin{equation}
d \sigma[A+B \to H + X] = \sum_n d \sigma[A+B \to (Q \bar Q)_n + X]~
\langle  {\cal O}_n^{H} \rangle.
\label{NRQCD-fact}
\end{equation}
The sum over $n$ includes the color, spin, and orbital-angular-momentum 
states of the $Q \bar Q$ pair. 
The pQCD cross sections are
essentially inclusive {\it partonic cross sections} for creating the $Q \bar Q$ pair,
which can be expanded in powers of $\alpha_s(m_Q)$,   
convolved with {\it parton distributions} for the colliding hadrons $A$ and $B$. 
The NRQCD matrix element
$\langle{\cal O}_n^{H} \rangle$ is essentially the probability for a
$Q\bar Q$ pair created in the state $n$ to evolve into a final state 
that includes the quarkonium $H$.
It is a nonperturbative constant that scales with a definite power 
of the typical relative velocity $v$ of the $Q \bar Q$ pair in $H$. 
It can be expressed as the vacuum expectation value of a
four-fermion operator in NRQCD  \cite{Bodwin:1994jh}.
The operators in color-singlet matrix elements are local operators,
but those in color-octet matrix elements include Wilson lines \cite{Nayak:2005rw}. 
The color-singlet matrix element that is leading order in $v$ can be determined 
phenomenologically from an electromagnetic annihilation decay rate.
The color-octet matrix elements can only be determined
phenomenologically from measurements of quarkonium production.

Since the NRQCD matrix elements $\langle{\cal O}_n^{H} \rangle$
scale with definite powers of $v$ that depend on $n$, the sum over $n$
in Eq.~\ref{NRQCD-fact} can be interpreted as an expansion in powers of $v$. 
The predictive power of NRQCD factorization comes from truncating that
expansion.  The truncation in $v$ is more accurate
for bottomonium than for charmonium, since $v^2$ is smaller by  a factor of about 1/3. 
Since the relative velocity of the $Q \bar Q$ pair in an excited quarkonium state
is not as small as in the ground state,
the $v$ expansion of NRQCD may converge more slowly for the excited states.
Thus the truncation in $v$ may introduce larger errors for $\psi(2S)$ than for $J/\psi$.
By truncating in $v$ and using approximate symmetries of NRQCD, 
the number of nonperturbative constants can be reduced to just a few
for each orbital-angular-momentum multiplet of quarkonium. 
For a spin-triplet $S$-wave quarkonium state $H$, such as the $J/\psi$ or the $\Upsilon(1S)$, 
the leading NRQCD matrix element is a color-singlet matrix element 
of order $v^3$ denoted by $\langle{\cal O}^{H}(^3S_1^{[1]})\rangle$.
It can be determined phenomenologically from the decay rate of $H$ into a lepton pair.
The truncation for S-wave states that is used in current phenomenology 
includes the NRQCD matrix elements through relative order $v^4$.  
There are three independent color-octet matrix elements denoted by
$\langle{\cal O}^{H}(^1S_0^{[8]})\rangle$,
$\langle{\cal O}^{H}(^3S_1^{[8]})\rangle$, and
$\langle{\cal O}^{H}(^3P_0^{[8]})\rangle$,
which are suppressed by orders $v^3$, $v^4$, and $v^4$, respectively.  The symbols 
in parentheses indicate the angular-momentum state $^{2S+1}L_J$ of the $Q\bar Q$ pair 
and whether its color state is singlet $[1]$ or octet  $[8]$. 
The truncation of the velocity expansion of NRQCD could be extended 
to a higher order in $v$ only
at the expense of introducing several additional phenomenological parameters.

When $p_T > m_Q$,
the NRQCD factorization formula can be expanded in powers of $m_Q/p_T$.
In the production of a spin-triplet $S$-wave quarkonium state,
the cross sections in the various NRQCD channels have 
different behaviors at large $p_T$.  At LO (leading order) in $\alpha_s$, 
which is order $\alpha_s^3(m_Q)$, 
the only channel at LP (leading power) is ${}^3S_1^{[8]}$.
The other color-octet channels ${}^1S_0^{[8]}$ and
${}^3P_0^{[8]}$ are NLP, and the color-singlet channel ${}^3S_1^{[1]}$
is N$^2$LP.
At NLO in $\alpha_s$, all three color-octet channels are LP, while
the color-singlet channel is NLP.
The suppression of the color-singlet channel by powers of $\alpha_s$ 
and $m_Q/p_T$ makes the color-octet channels important, 
despite their suppression by powers of $v$.

The NRQCD factorization formula is predictive of the quarkonium polarization.
With the truncation for $S$-waves at relative order $v^4$,
the polarization is determined by the same four NRQCD matrix elements
as the cross sections summed over quarkonium spins.
In principle, measurements of hadroproduction cross sections 
summed over quarkonium spins 
could be used to determine the matrix elements and then predict the polarization.
However, the $^1S_0^{[8]}$ and
$^3P_0^{[8]}$ terms in the hadroproduction cross sections 
have similar dependence on kinematical variables, such as $p_T$.  
Thus, to determine them separately,
one must in practice either use data from other production processes 
or else use polarization data.

\subsubsection{LP Fragmentation Formula}
\label{sec:LP-frag}

The LP fragmentation formula is a rigorous factorization formula 
in which nonperturbative effects associated with the binding of a $Q \bar Q$ pair 
into quarkonium are organized into functions, 
instead of multiplicative constants as in Eq.~\ref{NRQCD-fact}.
It states that the leading power (LP) of $m_Q/p_T$ 
in the cross section for producing quarkonium at large $p_T$
can be expressed as a sum of inclusive pQCD cross sections 
for producing a parton convolved with {\it fragmentation functions}:
\begin{equation}
d \sigma[A+B \to H + X] = \sum_i d\hat\sigma[A+B\to i+X]\otimes D_{i\to H}(z).
\label{LP-frag}
\end{equation}
The sum over $i$ extends over the types of partons (gluons, quarks, and antiquarks). 
The momentum of the parton $i$ is determined by the condition that the quarkonium $H$
has  longitudinal momentum fraction $z$ relative to the parton.
The ``$\otimes$'' in Eq.~\ref{LP-frag} represents an integral over $z$.
The pQCD cross sections  are
essentially inclusive partonic cross sections for producing the 
parton $i$, which can be expanded in powers of $\alpha_s(p_T)$,   
convolved with parton distributions for the colliding hadrons $A$ and $B$. 
The  fragmentation function $D_{i\to H}(z)$ is
the nonperturbative probability distribution for the momentum fraction $z$. 
The evolution equations for the fragmentation
functions can be used to sum large logarithms of $p_T/m_Q$ to all
orders in $\alpha_s$.
A proof of the LP factorization formula in Eq.~(\ref{LP-frag})  
was first sketched by Nayak, Qiu, and Sterman in 2005 \cite{Nayak:2005rt}. 

The LP fragmentation formula lacks the predictive power of the
NRQCD factorization formula, because the fragmentation functions 
$D_{i\to H}(z)$ are nonperturbative functions of $z$ that
must be determined phenomenologically.
Predictive power can be achieved
by applying the NRQCD factorization conjecture to the fragmentation functions.
It states that the fragmentation function for the parton $i$ to produce 
the quarkonium $H$ can be expressed as a sum of 
pQCD fragmentation functions multiplied by NRQCD matrix elements:
\begin{equation}
D_{i\to H}(z) = \sum_n d_{i \to (Q \bar Q)_n}(z)~
\langle  {\cal O}_n^{H} \rangle.
\label{D-fact}
\end{equation}
The pQCD fragmentation functions $d_{i \to (Q \bar Q)_n}(z)$
can be expanded in powers of $\alpha_s(m_Q)$.
The NRQCD-expanded LP fragmentation formula obtained by inserting 
Eq.~\ref{D-fact} into Eq.~\ref{LP-frag}
should reproduce the leading power in the expansion of  the
NRQCD factorization cross section in Eq.~\ref{NRQCD-fact}
in powers of $m_Q/p_T$.
The LP factorization formula was actually first applied to 
quarkonium production at large $p_T$ back in 1993, 
when the first fragmentation functions for quarkonium
were calculated to LO in $\alpha_s$ \cite{Braaten:1993rw,Braaten:1993mp}.
The fragmentation functions have since been calculated to NLO
for all the phenomenologically relevant channels and to
NLO for the $^3S_1^{[8]}$ channel \cite{Braaten:2000pc}. 
The usefulness of the NRQCD-expanded LP fragmentation formula 
has proved to be limited.
Explicit calculations using the NRQCD factorization formula have revealed 
that, in some channels, the LP cross section is
not the largest contribution until $p_T$ is almost an order of
magnitude larger than $m_Q$.

In the NRQCD-expanded LP fragmentation formula,
there are three expansion parameters: $\alpha_s(p_T)$, $\alpha_s(m_Q)$, and $v$.
The various NRQCD channels enter at different orders in $\alpha_s(m_Q)$.
For spin-triplet $S$-wave quarkonium,
the only channel that is LO in $\alpha_s$ is
${}^3S_1^{[8]}$ at order $\alpha_s^2(p_T) \alpha_s(m_Q)$.
The other color-octet channels ${}^1S_0^{[8]}$ and
${}^3P_0^{[8]}$ are NLO at order $\alpha_s^2(p_T) \alpha_s^2(m_Q)$.
The color-singlet channel ${}^3S_1^{[1]}$
is N$^2$LO at order $\alpha_s^2(p_T) \alpha_s^3(m_Q)$.
The suppression of the color-singlet channel by powers of $\alpha_s(m_Q)$
makes the color-octet channels important, 
despite their suppression by powers of $v$.

The NRQCD-expanded LP fragmentation formula has important implications for 
the polarization of spin-triplet $S$-wave quarkonium at large $p_T$.
The contribution that is LO in $\alpha_s(p_T)$ 
comes from production of a hard gluon.
At LO in $\alpha_s(m_c)$,  that gluon fragments into a $Q \bar Q$ pair 
in the ${}^3S_1^{[8]}$ channel \cite{Braaten:1994vv}.
The gluon is transversely polarized in the cm-helicity frame, and
at leading order in $v$, that polarization is transferred to the quarkonium.
Thus, at asymptotically large $p_T$, spin-triplet $S$-wave quarkonium
should be increasingly transversely polarized \cite{Cho:1994ih}.

\subsubsection{NLP Fragmentation Formula}
\label{sec:NLP-frag}

The NLP fragmentation formula is a rigorous extension of the
LP  fragmentation formula in Eq.~\ref{LP-frag} to the next-to-leading
power (NLP) of $m_Q^2/p_T^2$. Kang, Qiu, and Sterman proved in 2011 that the
terms suppressed by $m_Q^2/p_T^2$ can be written as
a sum of pQCD cross sections for producing a collinear $Q \bar Q$ pair
convolved with {\it double-parton fragmentation functions}
\cite{Kang:2011zza,Kang:2011mg}:
\begin{equation}
\sum_n d\hat\sigma[A+B\to (Q\bar Q)_n +X] \otimes D_{(Q\bar Q)_n \to H}(z, \zeta,\zeta').
\label{NLP-frag}
\end{equation}
The sum over $n$ extends over the color (singlet and octet) and Lorentz
(vector, axial-vector, and tensor) structures of the $Q \bar Q$ pair.
The pQCD cross sections are essentially
inclusive partonic cross sections for producing a collinear $Q \bar Q$
pair, which can be expanded in powers of $\alpha_s(p_T)$, 
convolved with parton distributions for the colliding hadrons $A$ and $B$. 
The double-parton fragmentation functions 
$D_{(Q\bar Q)_n \to H}(z, \zeta,\zeta')$  are nonperturbative probability
distributions in the longitudinal momentum fraction $z$ of the
quarkonium $H$ relative to the $Q \bar Q$ pair that 
also depend on the relative longitudinal
momentum fractions $\zeta$ and $\zeta'$ of the $Q$ and the $\bar Q$. 
The ``$\otimes$'' in Eq.~\ref{LP-frag} represents integrals over $z$, $\zeta$, and $\zeta'$.
The NLP fragmentation formula is obtained by adding Eq.~(\ref{NLP-frag}) 
to Eq.~(\ref{LP-frag}). The evolution equations for the fragmentation
functions can be used to sum large logarithms of $p_T/m_Q$ to all
orders in $\alpha_s$. 
A similar factorization formula has been derived by Fleming {\it et al.}\ using 
soft collinear effective theory \cite{Fleming:2012wy,Fleming:2013qu}, 
but it is not identical.  In particular, the form of the evolution equations for the 
fragmentation functions is different in the two approaches.

The NLP fragmentation formula lacks predictive power, 
because the double-parton fragmentation functions 
are nonperturbative functions of $z$, $\zeta$, and $\zeta'$
that must be determined phenomenologically. 
Predictive power can be achieved by applying the NRQCD factorization conjecture 
to the fragmentation functions.
The double-parton fragmentation functions $D_{(Q\bar Q)_n \to H}(z, \zeta,\zeta')$
have expansions in terms of the NRQCD matrix elements analogous to that for
the single-parton fragmentation function $D_{i\to H}(z)$ 
in Eq.~\ref{D-fact} \cite{Ma:2013yla,Ma:2014eja}.
The NRQCD-expanded NLP fragmentation formula
should reproduce the power expansion of the
NRQCD factorization cross section in Eq.~\ref{NRQCD-fact} 
up to order $m_Q^2/p_T^2$.

Kang, Qiu, and Sterman have taken the first step towards analyzing 
the effects of $Q \bar Q$ fragmentation on the polarization of 
quarkonium \cite{Kang:2011mg}.
For a spin-triplet $S$-wave quarkonium state,
the only $Q \bar Q$ fragmentation function that is nonzero
at LO in $\alpha_s$ is the color-octet axial-vector fragmentation function.
Its contribution to the cross section is increasingly longitudinal 
in the cm-helicity frame as $p_T$ increases.
They argued that the observed polarization of quarkonium could arise 
from a competition between a transversely polarized contribution 
from gluon fragmentation and a longitudinally polarized contribution 
from $Q \bar Q$ fragmentation.
At very large $p_T$, the leading power correction to the polarization comes from
interference between gluon fragmentation and $Q \bar Q$ fragmentation, 
and it falls like a single power of $m_Q/p_T$ \cite{Kang:2014tta}.
Thus NLP factorization predicts that the polarization at  
large $p_T$ should eventually be increasingly transverse, as predicted by
LP factorization, but this asymptotic behavior may not appear until very large $p_T$.

\subsubsection{Color-Singlet Model}
\label{sec:CSmodel}

One of the earliest attempts to describe quarkonium production 
using perturbative QCD was the {\it color-singlet model}
\cite{Kartvelishvili:1978id,Chang:1979nn,Berger:1980ni,Baier:1981uk}. 
A $Q \bar Q$ pair that is created in a high energy collision
is assumed to be able to bind to form a quarkonium $H$ 
only if it is created in a color-singlet state and in the same spin 
and orbital-angular-momentum state as the $Q \bar Q$ pair in $H$.
The color-singlet model can be obtained from the NRQCD factorization 
formula by assuming that the only nonzero NRQCD matrix element 
is the color-singlet matrix element that is leading order in $v$.
For a spin-triplet $S$-wave state $H$, such as the $J/\psi$ or the $\Upsilon(1S)$, 
that matrix element is  $\langle{\cal O}^{H}(^3S_1^{[1]})\rangle$.
Since this matrix element is determined by the decay rate of $H$ into a lepton pair,
the color-singlet model has no adjustable parameters.
In the  case of $P$-wave states, the color-singlet model is inconsistent,
because of infrared divergences at low orders in $\alpha_s$.

The color-singlet model gives unambiguous predictions for the polarization of 
spin-triplet $S$-wave quarkonium.
The predictions at LO and NLO in $\alpha_s$ are completely different
\cite{Artoisenet:2008fc,Gong:2008sn,Lansberg:2010vq,Butenschoen:2012px}.
At LO, the polarization in the cm-helicity frame is strongly transverse,
with $\lambda_\theta$ approaching 1 as $p_T$ increases,
while the polarization in the CS frame is  weakly longitudinal
and varies slowly with $p_T$.
At NLO, the polarization in the cm-helicity frame is increasingly longitudinal as $p_T$ increases,
while the polarization in the CS frame is  transverse and varies slowly with $p_T$.

\subsubsection{Color-Evaporation Model}
\label{sec:CEmodel}

The earliest attempt to describe quarkonium production using perturbative QCD was the 
{\it color-evaporation model} \cite{Fritzsch:1977ay,Halzen:1977rs}. 
A $Q \bar Q$ pair that is created in a high energy collision of hadrons
is assumed to be able to bind to form the quarkonium $H$ 
only if its invariant mass is below the open-heavy-flavor threshold.
The probability $f_H$ of binding is assumed to be independent of 
the color or spin state of the $Q \bar Q$ pair. 
If the $Q \bar Q$ phase space integrals are expanded around the threshold, 
the color-evaporation model reduces to the 
NRQCD factorization formula with simplifying assumptions 
about the NRQCD matrix elements \cite{Bodwin:2005hm}.

The color-evaporation model, as originally conceived, predicts zero polarization for 
quarkonium states.  
In principle, the model could be extended to give nontrivial predictions for polarization
by identifying the total spin of the $Q$ and $\bar Q$ with the spin of the quarkonium. 
Such an extension is not feasible in practice, because the cross sections 
in the color-evaporation model are calculated using NLO pQCD cross sections 
for producing $Q$ and $\bar Q$ in which their spin states 
have been summed over.

\subsubsection{$k_T$ Factorization}
\label{sec:kTfact}

The $k_T$-factorization approach is an alternative to standard collinear
factorization in which pQCD cross sections are expressed in terms of parton
distributions that depend on the transverse momenta of the partons, as
well as on their longitudinal momentum fractions. 
The $k_T$-factorization approach includes some contributions at leading
order in $\alpha_s$ that would appear only at higher orders in collinear
factorization.                  
The $k_T$-dependent parton distributions are known phenomenologically 
with much less precision than the collinear parton distributions.
The transverse momentum of the colliding partons is not expected 
to be important at large $p_T$.

In the applications of $k_T$ factorization to quarkonium production, 
production is usually assumed to occur only through 
the color-singlet $Q\bar Q$ channel that is leading order  in $v$
and the pQCD cross sections are usually calculated only to LO in $\alpha_s$. 
In this approximation, the polarization of the $\Upsilon(1S)$ 
in the cm-helicity frame is predicted to be increasingly longitudinal 
as $p_T$ increases  \cite{Baranov:2007ay,Baranov:2008yk}.

\subsection{NRQCD Factorization Phenomenology}
\label{sec:phenopolzn}

In order to use NRQCD factorization to predict the polarization of quarkonium,
the color-octet NRQCD matrix elements must be determined phenomenologically. 
Early predictions of the polarization were based on fits using pQCD cross sections 
calculated to LO  \cite{Braaten:1999qk,Kniehl:2000nn,Braaten:2000gw}.
Since then, three independent groups have carried out the heroic 
calculations of all the relevant pQCD cross sections  to NLO
\cite{Butenschoen:2010rq,Ma:2010jj,Gong:2012ug}.
Recent predictions of quarkonium polarization have been based on fits of the color-octet 
matrix elements using NLO pQCD cross sections.

In order to predict the polarization, it is essential to take into account the feeddown
from the direct production of higher quarkonium states.
The prompt production rate for $J/\psi$ 
includes significant feeddown from the direct production of $\psi(2S)$ and $\chi_{cJ}(1P)$.
The prompt production rate for $\Upsilon(1S)$ includes significant 
feeddown from the direct production of $\Upsilon(2S)$, $\Upsilon(3S)$, 
$\chi_{bJ}(1P)$, and $\chi_{bJ}(2P)$.  
The feeddown contributions to the unpolarized cross sections for $J/\psi$ 
and $\Upsilon(1S)$ are about $30$ or $40\%$.
However, the feeddown contributions could have a larger effect on the polarization.

NRQCD predictions for the polarization of the $J/\psi$ vary dramatically, 
depending on the data used to determine the color-octet NRQCD matrix elements.
All of the groups include the data from CDF Run~II for $d\sigma/dp_T$ 
with $p_T$ greater than $7$~GeV.
A prediction of strong transverse polarization in the cm-helicity frame \cite{Butenschoen:2012px} 
arises if one includes in the fits the HERA data for photoproduction of $J/\psi$
down to a $p_T$ of $3$~GeV. 
A prediction of moderate transverse polarization \cite{Gong:2012ug} 
arises if one includes in the fits the data from LHCb for $d\sigma/dp_T$
with $p_T$ greater than $7$~GeV 
and if one uses NRQCD factorization predictions to
correct for feeddown from the $\psi(2S)$ and the $\chi_{cJ}$ states. 
A prediction of near-zero transverse polarization \cite{Chao:2012iv}
arises if one includes in the fits the CDF Run~II polarization measurement. 

A complete NLO NRQCD analysis of the $\Upsilon(nS)$ states,
including the effects of feeddown, has recently been carried out \cite{Gong:2013qka}.
The color-octet matrix elements for the $S$-wave and $P$-wave states
were determined by fitting cross sections and polarization data 
measured at the Tevatron and the LHC with $p_T > 8$~GeV.
The $\Upsilon(1S)$ and $\Upsilon(2S)$ states are predicted to have 
small transverse polarizations in the cm-helicity frame.
The  $\Upsilon(3S)$ is predicted to have a more rapidly increasing 
transverse polarization as $p_T$ increases.
The difference might not be as dramatic if feeddown 
from the $\chi_{bJ}(3P)$ were taken into account.

\section{EXPERIMENTAL ISSUES for POLARIZATION MEASUREMENTS}

Hadronic production processes for quarkonium states contain two 
{\it prompt} contributions in which the quarkonium is produced at 
the hadronic collision point:
\begin{itemize}
\item[(a)]
{\it direct} production, in which the quarkonium is produced by
the binding of a heavy quark and antiquark created by the 
strong interactions of QCD,  
\item[(b)]
{\it feeddown}, in which the quarkonium is produced by hadronic
or electromagnetic transitions from a higher state in the 
quarkonium spectrum that was produced directly.  
\end{itemize}
At large center-of-mass energy $\sqrt{s}$, $B$-hadron decays produce 
additional charmonium 
events that are removed by comparing the location of the dilepton vertex  
with that of the primary vertex where the hadrons collided.

For the spin-triplet $S$-wave states of interest, feeddown events
come from decays of higher radial excitations and orbital-angular-momentum 
 excitations.  This can modify both the polarization and 
the kinematic variables of the lower-mass $S$-wave state compared 
to its direct production properties.  The photons and pions from the 
feeddown transitions have low energy.  In a hadronic production 
environment, it is difficult to 
measure such low-energy tracks and to associate them correctly with 
the dimuon pair in order to separate feeddown and direct production. 

The material for the bulk of this review 
originates from experiments done with $p \overline{p}$ collisions 
at 1.8 and 1.96 TeV center-of-mass energy at the Fermilab Tevatron 
and with $pp$ collisions at 7 TeV center-of-mass energy 
at the Large Hadron Collider (LHC) at CERN.  Other studies at lower 
$\sqrt{s}$, including $J/\psi$ and $\Upsilon$ polarization studies in
 $pA$ collisions 
by the E866 (NuSea) Collaboration at Fermilab, $J/\psi$ 
polarization measurements in $pA$ collisions by the HERA-B Collaboration 
at DESY, and $J/\psi$ polarization measurements in $pp$ collisions 
by the PHENIX Collaboration at Brookhaven, contribute results at smaller $p_T$.

Measuring quarkonium polarization puts stringent requirements on experiment 
design and apparatus performance.  Polarization is a differential 
measurement; it uses the lab frame trajectories of two leptons to 
determine the decay momentum vector of $\ell^+$ with respect to a 
quantization axis in the dilepton rest frame.  For polarization, not only 
must one know the apparatus efficiency for all events within the acceptance 
coverage, but also the acceptance must cover a large fraction of the decay 
angular variables in order to determine the three polarization 
parameters $\lambda_\theta$, $\lambda_\phi$, and $\lambda_{\theta \phi}$
that describe the decay process (Eq.~\ref{dW-dtheta,dphi}).  In 
polarization experiments,  
the more complete the angular phase space coverage, 
the better the determination of the polarization parameters.  
Until recently, quarkonium polarization measurements have focussed 
on just the single parameter $\lambda_{\theta}$ 
for a specific spin-quantization axis.  
Modern experiments, prompted by the discussions in 
Ref.~\cite{fac1,fac3}, have moved to measure all 
three polarization parameters in several reference frames in their 
analyses.  One important advantage is that this 
allows the frame-invariant polarization parameters
$\tilde{\lambda}$ and ${\tilde \lambda}'$ defined in Sec.~\ref{frames} to be used 
as diagnostics for possible inconsistencies.

\subsection{Background Determination and Angular Characteristics}

Polarization experiments are sensitive to the angular structure of 
the background in the dilepton rest frame.  Demonstrating good 
control of the background angular distribution is essential for any 
polarization experiment.  In almost all experiments, the background 
definition procedure is the same: 
\begin{enumerate}
\item
make a dilepton mass plot of selected prompt or 
(for charmonium only) $B$-decay events in each analysis phase space bin 
($\Delta p_T$,$\Delta x_F$ or $\Delta y$,$\Delta \cos \theta$); 
\item
fit the sideband background to a suitable empirical function 
and the signal shape to a predetermined functional form, often 
based on simulation;
\item
either subtract the estimated background 
to determine the signal yield and uncertainty in that phase space bin 
or, when sample sizes are large, 
make a simultaneous fit to signal distributions and background 
distributions to maximize the statistical power.  
\end{enumerate}
For experiments that analyze the two-dimensional decay angular 
distribution, the phase space becomes four-dimensional; binning in 
the azimuthal angle $\phi$ is also required, and the mass plot is 
done in $\cos \theta$--$\phi$ space for each  
bin of $p_T$ and the longitudinal variable.  

\subsection{Apparatus Acceptance} 

Every polarization measurement must determine the apparatus acceptance
after all the kinematic selections on the individual 
leptons have been applied and also must determine the efficiency for 
triggering on and 
detecting each of the leptons that falls into the acceptance.  
Simulation techniques 
are primary tools for these studies, but experimental validation of 
the Monte Carlo results is highly desirable.    
Using GEANT-based simulation models~\cite{geant} makes the acceptance
 calculations robust.  Single-lepton kinematic distributions for data
 and Monte Carlo samples are compared to validate the quarkonium 
kinematic parameters of the simulation.  Two different approaches are used: 
\begin{itemize}
\item[(a)]
generate {\it only} quarkonium events using an 
event generator, typically EvtGen~\cite{EvtGen}, that specifies the 
laboratory frame kinematics ($p_T$ and $x_F$ or $|y|$ distributions) 
of the quarkonium based on other measurements, perhaps by the same 
group, 
\item[(b)]
use a complete event generator like Pythia~\cite{Pythia} to 
generate the quarkonium states inclusively, with kinematic 
distributions 
chosen by Pythia or possibly adjusted for the experiment.  
\end{itemize}
In order to use option (a), the experiment has to have 
a tracking detector with low average occupancy, so that lepton track 
reconstruction is unlikely 
to be distorted by the presence of other tracks in the event.  
 For option (b) with Pythia generation, 
the simulation procedure generates the quarkonium 
kinematics and also produces the decay.  If reweighting has to be 
done to match kinematic parameters, it occurs {\it ex post facto}, 
and there can be distortion of the generated event 
distribution in the quarkonium rest frame due to reweighting. 
The quantization axis of the generated event is not the same as the 
corresponding quantity after reweighting.  Of course, the extent of 
any such shift can be studied and its impact on the polarization 
parameters evaluated within the Monte Carlo framework.

The lepton detection efficiency can vary with the kinematic variables
of the lepton $\ell$, usually its transverse momentum $p_T(\ell)$ 
with respect to the beam direction and its pseudorapidity 
$\eta(\ell) = - \ln[\tan(\theta/2)]$.  
The tag-and-probe method of 
efficiency determination is used in most modern 
experiments~\cite{cdf_run1_jpol}.  The {\it probe} track, unbiassed by
 a trigger, is either passed or failed by the analysis criteria.  
Probe track efficiencies in ($p_T, \eta$) bins can give the 
distribution of the trigger and detection efficiency for a single 
track as a function of its 
kinematic parameters.  The tag-and-probe method can also be applied 
to Monte Carlo 
samples with large statistics.  If the simulation and data efficiency
distributions agree, then one can use the Monte Carlo shape to fit to
 the data in order to improve knowledge of the kinematic variation 
of efficiencies.  
In some experiments, tag-and-probe studies are not possible, so  
simulation studies provide both efficiency 
and acceptance.  This introduces a level of uncertainty into the 
results that can be hard to quantify.

\section{EXPERIMENTS and RESULTS}

The general procedures outlined in the previous section have been used 
both in fixed target experiments (using proton or pion beams) and in 
collider experiments at the Tevatron ($p \overline{p}$ collisions) 
and at Fermilab, RHIC or LHC ($p p$ or $pA$ collisions).  We focus 
here on the experiments that have produced the highest statistics 
measurements for charmonium and bottomonium polarization 
in each class of experiments.

\subsection{ Fixed Target Experiments}

Two very different fixed target experiments dominate the field: 
Fermilab E866 (NuSea)~\cite{nusea_jpol, nusea_bpol} at 
$\sqrt{s} = 38.8$~GeV and HERA-B~\cite{herab_jpol} at 
$\sqrt{s} = 41.6$~GeV.  Both use nuclear targets.  At these center-of-mass 
energies, $\psi(2S)$ production and non-prompt $J/\psi$ production are
 negligible.  The transverse momentum $p_T$ is, at best, comparable 
to the charmonium mass.  
Because these are well-discussed experiments, we only summarize their
 results.  For NuSea, the polarization  parameter $\lambda_{\theta}$ 
in the Collins-Soper (CS) frame for $J/\psi$ polarization is small over 
the $p_T$ range $< 4$~GeV, with an average of +0.15 for 
$\langle x_F \rangle = 0.45$.  
For HERA-B, the effective $\lambda_{\theta}$ parameter
in the CS frame is negative, averaging to $-0.18$ for 
$\langle x_F \rangle = -0.12$.  
This may indicate sensitivity of the polarization 
 to the production $x_F$ range 
at low $p_T$ in $pA$ collisions.  The HERA-B results 
confirm that the three polarization parameters are all small, both 
in the cm-helicity and CS frames.

The NuSea $\Upsilon$ analysis finds that the combined system for 
$\Upsilon(2S)$ and $\Upsilon(3S)$ has a polarization very similar 
to that of 
Drell-Yan dimuon pairs in the Collins-Soper frame~\cite{DYar}, 
fully transversely 
polarized for $p_T< 4$~GeV for $\langle x_F \rangle \sim 0.23$.  
In contrast, 
the $\Upsilon(1S)$ $\lambda_{\theta}$ parameter is essentially 
zero for $p_T < 1.8$~GeV.  

\subsection{Tevatron Polarization Measurements}
At collider energies, one has to consider both prompt and decay 
sources for charmonium.  For colliding beams, $x_F$ will always be 
small because $p_{\rm max}$ is large.   The appropriate kinematic 
variables for quarkonium are transverse momentum $p_T$ and rapidity 
$y$. The first Tevatron collider measurements were made in the 
central region $|y| < 0.6$ with $p_T < 20$ GeV.  
The CDF collaboration reported that the fraction of $J/\psi$ mesons 
that came from $\chi_c$ feeddown was $0.45 \pm 0.05 \pm 0.15$ 
for $p_T > 6$~GeV, $|y|<0.5$ \cite{cdf_xs2}.  The 
D0 collaboration reported that this feeddown fraction was 
$0.35 \pm 0.07 \pm 0.07$ for $p_T > 8$~GeV, $|y|<0.6$ \cite{d0_xs1}.  
For bottomonium, CDF~\cite{cdf_xs3} determined that the 
fraction of directly-produced 
$\Upsilon(1S)$ mesons having $p_T > 8$~GeV is $0.509 \pm 0.082 \pm 0.090$, 
very similar (the same within uncertainties) to the $J/\psi$ result.  
The feeddown fraction in quarkonium production seems 
to be at most mildly dependent on beam energy, $p_T$ range, 
target type, and which heavy quark is involved.  
We will want to look at polarization systematics in this light.

In Run~1 of the Tevatron, CDF made measurements of $J/\psi$, 
$\psi(2S)$, and $\Upsilon(1S)$ polarization at $\sqrt{s} = 1.8$~TeV. 
In Run~2 of the Tevatron, $\sqrt{s}$ was increased to 1.96~TeV and
the integrated luminosity increased by more 
than an order of magnitude.  
Both CDF and D0 made measurements of bottomonium 
polarization, and CDF repeated its study of $J/\psi$ and 
$\psi(2S)$ polarization.

\subsubsection{Quarkonium Polarization at $\sqrt{s} = 1.8$~TeV}

The CDF polarization measurements in the cm-helicity frame from Run~1
 are well known~\cite{cdf_run1_jpol}.  We note that in this 
pioneering experiment, covering a rapidity range $|y| < 0.6$, only 
60\% of the detected muons were measured in the vertex detector.  
This makes the determination of the efficiency more difficult for 
asymmetric decays compared to later experiments that had full 
coverage.   For prompt $J/\psi$ events, the average 
$\langle \lambda_{\theta} \rangle = +0.21 \pm 0.05$ and all 
measurements were positive for the range $p_T <  15$~GeV.  However, 
there was no suggestion that the polarization was becoming more 
transverse as $p_T$ increased.  The $\psi(2S)$ 
polarization was measured, but had large uncertainties.

For the $\Upsilon(nS)$ states, the yields of the higher excited states
 were low, and only the $\Upsilon(1S)$ polarization is 
reported~\cite{cdf_run1_ypol}.  The rapidity range was $|y| < 0.4$, 
so that 
the acceptance of the vertex detector was somewhat larger than for the 
$J/\psi$ case. The polarization parameter $\lambda_{\theta}$ 
 in the cm-helicity frame was measured in four $p_T$ bins.  All are 
consistent with zero within one standard deviation, with 
$\langle \lambda_{\theta} \rangle = -0.12 \pm 0.22$. 

\subsubsection{D0 Run~2 $\Upsilon(nS)$ Polarization}

The D0 Collaboration measured the $\Upsilon(nS)$ polarization 
in the cm-helicity frame~\cite{d0_ypol}.  The dataset covered a 
large rapidity 
range, $|y| < 1.8$.  High rapidity events have poorer mass resolution
 than central events, and the three $\Upsilon(nS)$ states overlapped 
in the mass distribution.  The D0 analysis imposed a muon isolation cut 
to purify the sample.  This has not been done in any other 
polarization experiment.  Unlike other experiments, there is a 
very small sideband region on the low mass side of the 
$\Upsilon(1S)$ signal region due to a combination of poor mass 
resolution and a dimuon trigger threshold. The background shape under
the broad signal region is poorly constrained by the data.  

The D0 simulation uses Pythia to study unpolarized 
$\Upsilon(1S)$ (or $\Upsilon(2S)$) decays to two muons. The dimuon $p_T$ 
distribution and the total momentum 
from the Monte Carlo were reweighted to match the data.    
This influences the helicity boost and a 
systematic uncertainty is assigned.  After corrections, the simulated
 $\Upsilon(1S)$ mass peak is 40~MeV different from the PDG value.  
This is a much larger discrepancy than is seen in the simulations 
from other experiments.  The measured $\lambda_{\theta}$ 
parameter as a function of $p_T$ for the $\Upsilon(1S)$ 
in the range $|y|< 1.8$ is very different from what was 
reported by CDF in the Run~1 measurement for a similar 
$p_T$ range but covering only the central rapidity region 
$|y|< 0.4$.  For $p_T < 10$~GeV, 
$\langle \lambda_{\theta} \rangle = -0.45 \pm 0.06$. 

\subsubsection{CDF Run~2 Charmonium Polarization}

The CDF Run~2 $J/\psi$ and $\psi(2S)$ polarization measurements were 
done in 
the cm-helicity frame for $|y|<$ 0.6~\cite{cdf_run2_jpol}.  All 
tracks with 
$\eta <$ 0.6 traversed the silicon vertex detector, improving the 
efficiency 
compared to Run~1.  The analysis followed the same methodology
as the Run~1 measurement.  A simultaneous fit to the dimuon mass and 
the 
transverse vertex position separated events into prompt and 
$B$-hadron decay candidates.  Muon efficiencies 
and trigger efficiencies were determined by experimental 
tag-and-probe studies for all three trigger levels.  
These efficiencies were applied to simulated muons in fully polarized
 (T or L) Monte Carlo samples in order to account for apparatus effects.  

The average $\lambda_{\theta}$ parameter in the cm-helicity frame 
for the $J/\psi$ from this analysis is small and consistently negative
  $\langle \lambda_{\theta} \rangle = -0.062 \pm 0.013$.  This disagrees
 with the CDF Run~1 result.  On the other hand, the $B$-hadron decay 
polarization for the Run~2 data gives an effective 
$\lambda_{\theta}^B = -0.11 \pm 0.04$, consistent both with 
the Run~1 result and with the Monte Carlo simulation of 
$B \rightarrow J/\psi X$ decays.  Large statistical uncertainties 
preclude any statement about the $\psi(2S)$ polarization.

\subsubsection{CDF Run~2 Bottomonium Polarization}

The CDF Run~2 study of $\Upsilon(nS)$ polarization~\cite{cdf_run2_ypol} 
for a rapidity range $|y| < 0.6$ introduced new analysis steps to 
improve background control 
and yielded several first-time results.  This study, with better 
statistical accuracy than any other measurement, is the first 
to make a simultaneous determination of the three 
polarization parameters for the two-dimensional $(\cos \theta,\phi)$ 
distribution using the methods discussed in Sec.~\ref{frames}.  It 
also produced 
the first measurement of the $\Upsilon(3S)$ polarization parameters.
The trigger efficiency and 
single-muon efficiencies were evaluated using tag-and-probe analyses 
for all three dimuon trigger levels.  The acceptance was determined 
from unpolarized Monte Carlo simulations.

The analysis for the three polarization parameters
$\lambda_\theta$, $\lambda_\phi$, and $\lambda_{\theta\phi}$
for each mass peak 
was done following the outline in Sec.~\ref{frames}.  It covers 
the range $2 < p_T < 40$~GeV.  The data were separated into $p_T$ bins.  
In each bin, the data were boosted to the cm-helicity or CS analysis frame 
and the angular variables $(\cos \theta, \phi)$ 
were divided into $0.05 \times 5^{\circ}$ bins.  
Ref.~\cite{cdf_run2_ypol} gives 
the details of how the displaced sample background was used 
to constrain the background in the signal region by a series of fits.
  Independent fits were done in the cm-helicity frame 
and the CS frame and their consistency is validated using 
the frame-invariant polarization parameter $\tilde{\lambda}$.  
There is no indication that there is 
any high-$p_T$ change in the polarization for any of the three states.  
These are the most precise determinations of the $\Upsilon(nS)$ now 
available.  For the $\Upsilon(1S)$ in the cm-helicity frame, 
$\langle \lambda_{\theta} \rangle = -0.102 \pm 0.027$ 
for $p_T < 12$~GeV, agreeing with CDF Run~1 and disagreeing with D0.

\subsection{RHIC and LHC Polarization Studies}
 
The advent of $pp$ colliders allowed quarkonium studies with targets 
similar to the fixed target studies, but in a very different 
kinematic regime.  PHENIX at RHIC and ALICE, CMS, and LHCb at LHC have 
published $J/\psi$ polarization studies and CMS has published results
 on $\Upsilon(nS)$ polarizations.  The ALICE results are dominated by
 those of LHCb, which covers the kinematic range 
$2  < p_T <  15$~GeV and forward rapidity $2 < y < 4.5$.  
The CMS results cover 
the high-$p_T$ (14-70~GeV), central rapidity ($|y| < 1.2$) regime.  
All three experiments present results in both the cm-helicity and CS frames. 
The random dimuon background for the quarkonium states is much lower 
at the LHC than for the Tevatron.  This simplifies the background subtraction.

\subsubsection{CMS Bottomonium Polarization}

Polarization results for all three $\Upsilon(nS)$ states having 
$10 < p_T < 50$~GeV and two rapidity ranges for $|y| < 1.2$ are 
reported by CMS in Ref.~\cite{cms_ypol}. 
The offline minimum $p_T$ requirement was 10~GeV to ensure stable 
efficiency measurements.  The single-muon trigger efficiencies were 
based on tag-and-probe studies.  

The analysis grouped dimuon events in $p_T$ bins for $|y| < 0.6$ 
or $0.6 < |y| < 1.2$.  The CMS measurement, like the CDF study, 
shows that all 
three polarization parameters are small for all three $\Upsilon$ 
states, both in the cm-helicity and the CS frames.  
The frame-invariant variable $\tilde{\lambda}$ 
for each state indicates good agreement between frames, 
compatible with the range of variation expected from simulation studies.   
The $\Upsilon(3S)$ polarization parameters do not rise at large $p_T$.
  For the $\Upsilon(1S)$ in the cm-helicity frame,  
$\langle \lambda_{\theta} \rangle = +0.074 \pm 0.064$ for the range $10
 < p_T < 30$~GeV for $pp$.

\subsubsection{CMS Charmonium Polarization}

The CMS Collaboration used the same analysis technique to determine 
the polarization parameters of charmonium~\cite{cms_jpol}.  
This data set has the largest reported sample of $J/\psi$ mesons 
used for polarization analysis and provides the first meaningful 
polarization measurements for the $\psi(2S)$.  
For the $J/\psi$, results are reported for $14 < p_T < 70$~GeV in 
two rapidity ranges: $|y| < 0.6$ and $0.6 < |y| < 1.2$.  For the 
$\psi(2S)$, 
the phase space covered was $14 < p_T < 50$~GeV in three rapidity 
ranges: $|y| < 0.6$, $0.6 < |y| < 1.2$, and $1.2 < |y| < 1.5$.

Events were separated into prompt and $B$-hadron decay candidates by 
making a fit to the transverse vertex displacement from the primary.
  There are no large polarization parameters in either the cm-helicity 
or CS frame, and the $\tilde{\lambda}$ test shows good consistency 
for the analysis. Detailed results will be discussed below.

\subsubsection{LHCb and ALICE Charmonium Polarization}

These two experiments report $J/\psi$ polarization parameters 
at forward rapidity in the cm-helicity and CS frames.  In 
ALICE~\cite{alice_jpol},
a muon spectrometer gave forward coverage with modest mass resolution.
  Contributions from $\psi(2S)$ were ignored.  Polarization 
variables $\lambda_{\theta}$ 
and $\lambda_{\phi}$ were determined from the dimuon mass spectrum 
binned in 
$\cos \theta$ or $\phi$.  The sidebands are used to subtract 
the background under the $J/\psi$ mass peak in each projected angle 
bin.  The resulting experimental distributions were corrected for 
efficiency and for acceptance using simulated events from an 
unpolarized Monte Carlo study.  

The LHCb~\cite{lhcb_jpol}
coverage was $2 < p_T < 15$~GeV for $2 < y < 4.5$ with good mass 
resolution.
The LHCb analysis also relies on simulation to determine efficiency
 and acceptance, but they have a major advantage in calibrating the
 results -- 
a sample of fully-reconstructed $B^+ \rightarrow J/\psi\,  K^+$ decays 
for which the $J/\psi$ polarization is known.    
The LHCb mass distribution for $J/\psi$ events has very little 
background.  The usual subtraction technique is used.  

Both LHCb and 
ALICE observe small polarization in this $p_T$ range, and the 
other two polarization parameters (just $\lambda_{\phi}$ for ALICE) 
are consistent with being zero in the 
cm-helicity frame throughout this $p_T$ range. The ALICE analysis finds 
$\langle \lambda_{\theta} \rangle = -0.14 \pm 0.10$.  
The average $\lambda_{\theta}$ parameter 
from LHCb is $-0.063 \pm 0.011$ for the range $2 < p_T < 15$~GeV.  
Overall the LHCb polarization 
parameter measurements in either frame show little variation with 
$y$ or $p_T$ within their joint ranges.

\subsubsection{PHENIX Charmonium Polarization}

The PHENIX Collaboration at RHIC identified $J/\psi$ events in the $e^+e^-$ 
channel from 200~GeV $pp$ collisions with $|y| < 0.35$ and 
$p_T(ee) < 5$~GeV.~\cite{phenix}.   Electron efficiencies from 
simulation were calibrated  
using photon conversion electrons from the beam pipe and unpolarized 
simulated decays were used to correct the data and determine  
$\lambda_{\theta}$ in the cm-helicity and Gottfried-Jackson (GJ) 
frames.   The PHENIX $pp$ 
phase space for $\sqrt{s}$=200~GeV is similar to that for the 
HERA-B $pA$ measurement at $\sqrt{s}$=41.6~GeV, and the PHENIX 
polarization results agree with the more precise HERA-B measurements.

\subsection{Experimental Summary}

What can we conclude from the variety of experimental results 
presented here?  We choose to present the measurements in two ranges:
 $p_T < 10$~GeV and $p_T > 10$~GeV.  This choice reflects both the 
experimental information and the $p_T$ behavior of the 
quarkonium production cross sections.  Roughly independent of target 
and $\sqrt{s}$, the differential cross section peaks near $p_T 
\sim 2$~GeV (4 GeV) for $J/\psi \ (\Upsilon(1S))$ production.  For 
$p_T > $ 10~GeV it falls smoothly for both states. The low $p_T$ 
range has data from collider and fixed 
target experiments.  The large $p_T$ range is covered only by 
collider experiments. A first question to consider is how much does 
$\sqrt{s}$ matter in polarization results at low $p_T$.  The 
experiments cover the range 38.8~GeV  $< \sqrt{s} <$ 7~TeV for $pA$ 
collisions, 
$p\overline{p}$ collisions, and $pp$ collisions at both low and 
high rapidity.

\subsubsection{Polarization Results for $J/\psi$ with $p_T < 10$~GeV}

In general, the $J/\psi$ polarization measurements in the cm-helicity
 or CS frames vary somewhat among experiments, 
but the polarization parameters are never 
large.  The contributing experiments are HERA-B 
with $pA$ collisions, ALICE, LHCb and PHENIX with $pp$ collisions, 
and CDF Run~1 and Run~2 with $p\overline{p}$ collisions.  
Of these, ALICE and LHCb are large rapidity measurements; 
the others are central.   

\begin{figure}
\centerline{ \includegraphics*[width=16cm,clip=true]{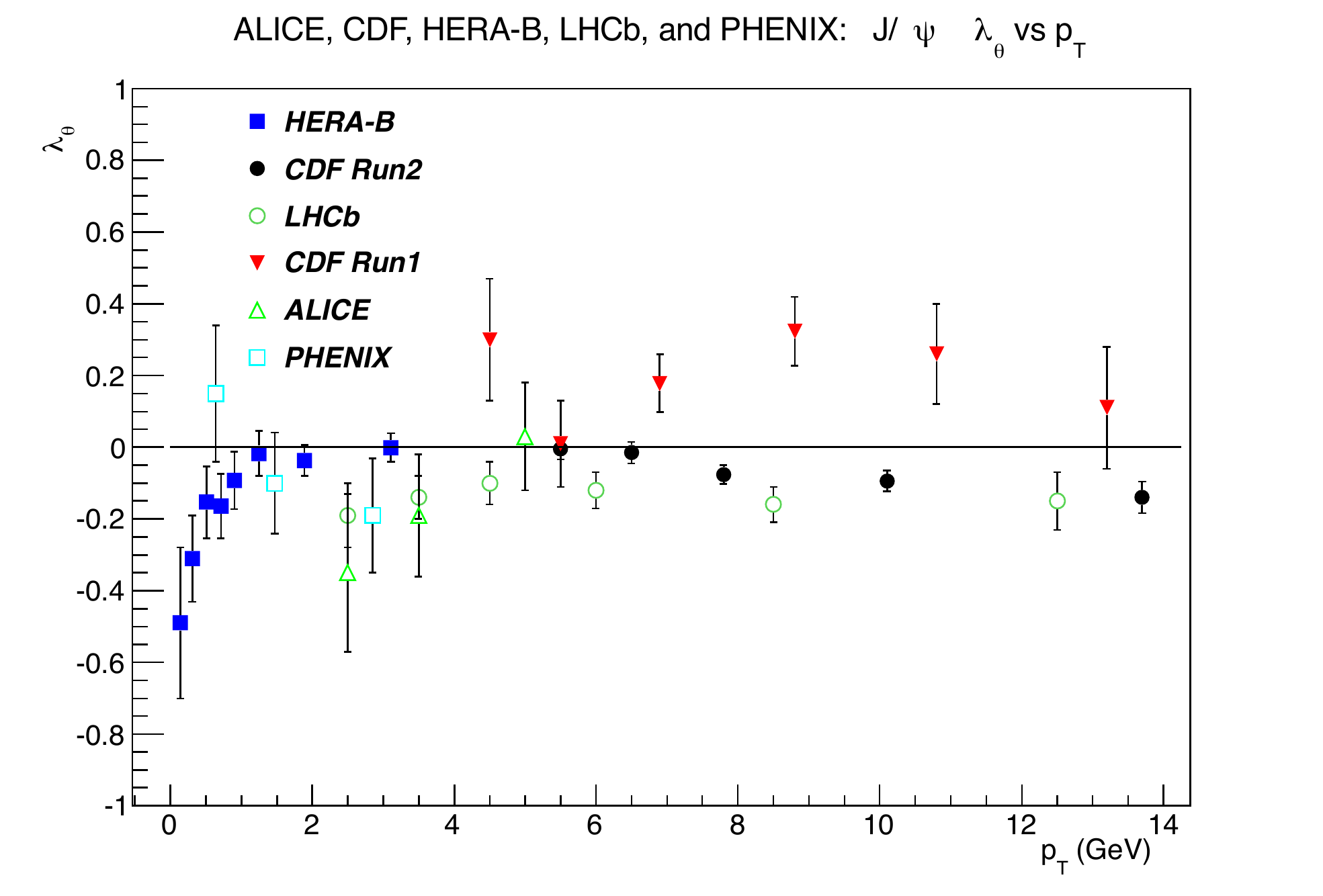}}
\caption{Measurements of $\lambda_{\theta}$ in the cm-helicity frame 
for $J/\psi$ production  with $p_T < 10$~GeV.  
The data are from  ALICE, CDF Run~1, CDF Run~2, HERA-B, LHCb, and PHENIX. 
\label{jpollow}}
\end{figure}

The $\lambda_{\theta}$ parameter in the cm-helicity frame for the 
six experiments is plotted versus $p_T$ in Fig.~\ref{jpollow}.  
The HERA-B $\lambda_{\theta}$ measurement is nearly zero for $p_T > 1$
~GeV.  For PHENIX over the range $0< p_T < 5$~GeV, 
$\langle \lambda_{\theta} \rangle = -0.10^{+0.05}_{-0.09}\pm 0.05$.  
The CDF Run~2 average is $\lambda_{\theta} = -0.035 \pm 0.016$ 
for $5 < p_T < 9$~GeV .  The CDF Run~1 average 
$\langle \lambda_{\theta} \rangle = +0.21 \pm 0.05$  disagrees with 
CDF Run~2. 
At low $p_T$ for $J/\psi$ production, ALICE, CDF Run~2, HERA-B, LHCb, 
and PHENIX agree that $\lambda_{\theta}$ in the cm-helicity frame 
is negative and close to zero for $p_T$ between 1 and 10~GeV 
independent of target, $\sqrt{s}$, or rapidity range.  
The CDF Run~1 result looks like an experimental outlier.

NuSea only reports data on $\lambda_{\theta}$ for forward $x_F$ 
in the CS frame, assuming $\lambda_{\phi}$ = 0.  That assumption 
is consistent with ALICE and HERA-B and LHCb observations.  

The overall picture shows that $J/\psi$ polarization for $p_T < 10$~GeV 
is small for production from any kind of target, at any $\sqrt{s}$, 
and any rapidity in both the cm-helicity and CS frames.  
In this $p_T$ range, no measurements of $\psi(2S)$ polarization 
give any useful limits.

\subsubsection{Polarization Results for $J/\psi$ and $\psi(2S)$  
with $p_T > 10$~GeV}

Data for $p_T > 10$~GeV come mostly from CMS measurements, 
which probe a new energy regime as well as extending the $p_T$ region.  
Feeddown effects are a complication for interpreting 
$J/\psi$ polarization results, so we look first at the $\psi(2S)$ 
results from CMS in the cm-helicity frame.
For $|y| < 0.6$, the $\psi(2S)$ polarization parameters are consistent
 with being $p_T$-independent in the range $14 < p_T <  50$~GeV.  
The average $\lambda_{\theta}$ for the interval is $0.13 \pm 0.12$, 
and gives no sign of becoming significantly transverse, even though 
uncertainties on individual points are not small.  At higher rapidity, 
$0.6 < |y| < 1.2$, the trend is again for a $p_T$--independent 
$\lambda_{\theta}$.  In the cm-helicity frame, 
$\langle \lambda_{\theta} \rangle = -0.092 \pm 0.088$, which is more negative 
than for $|y| <$ 0.6 but consistent to within 1.5 standard 
deviations. Polarization parameters for $\psi(2S)$ production are small 
both in the cm-helicity and CS frames and show little variation in the 
$(p_T,y)$ phase space of the CMS measurement.

The $J/\psi$ polarization from the CMS measurements again shows 
a stable $p_T$--independent pattern over the range 
$14 < p_T < 70$~GeV 
in both rapidity ranges for the cm-helicity frame.  Like the 
$\psi(2S)$ case, 
the $p_T$--averaged $\lambda_{\theta}$ value becomes slightly less 
positive at higher rapidity, with average values of $0.14 \pm 0.04$
 at smaller rapidity and $0.08 \pm 0.03$ at larger rapidity, but 
consistent within 1.5 standard deviations.  None of the three 
polarization parameters is large in either the cm-helicity or CS frame.

There is some tension between the LHCb results and the CMS results, 
even though there is no overlap in the data.  The large-rapidity LHCb
$\lambda_{\theta}$ results in the cm-helicity frame show a persistent 
negative polarization in the domain 
$p_T < 15$~GeV, $2 < y < 4.5$.  The CMS measurements for $|y| < 1.2$ and 
$p_T > 14$~GeV are all positive. The uncertainties on the 
individual bin measurements near the boundary regions in $(p_T,y)$ 
space are not small from either experiment, making it difficult 
to project a trend from one phase space region into the other.  
Future LHC measurements may clarify the issue.

\subsubsection{Polarization Results for $\Upsilon(nS)$ with $p_T < 10$~GeV}

Bottomonium polarization results at low $p_T$ have been published 
by NuSea, CDF Run~1, CDF Run~2, and D0 Run~2.  The NuSea results 
are in a different kinematic region from the other experiments 
and have no independent check.  For $0< x_F < 0.6$ and $p_T < 4$~GeV,
 the measured polarization in the CS frame from NuSea is transverse, 
like Drell-Yan polarization, for the two excited S-wave states 
$\Upsilon(2S)+\Upsilon(3S)$ 
compared to nearly zero polarization for the $\Upsilon(1S)$.  
Large polarization of any $\Upsilon(nS)$ state in the CS frame 
is not seen in high energy collider experiments.  
The pattern of the NuSea results is unusual.  

\begin{figure}
\centerline{ \includegraphics*[width=16cm,clip=true]{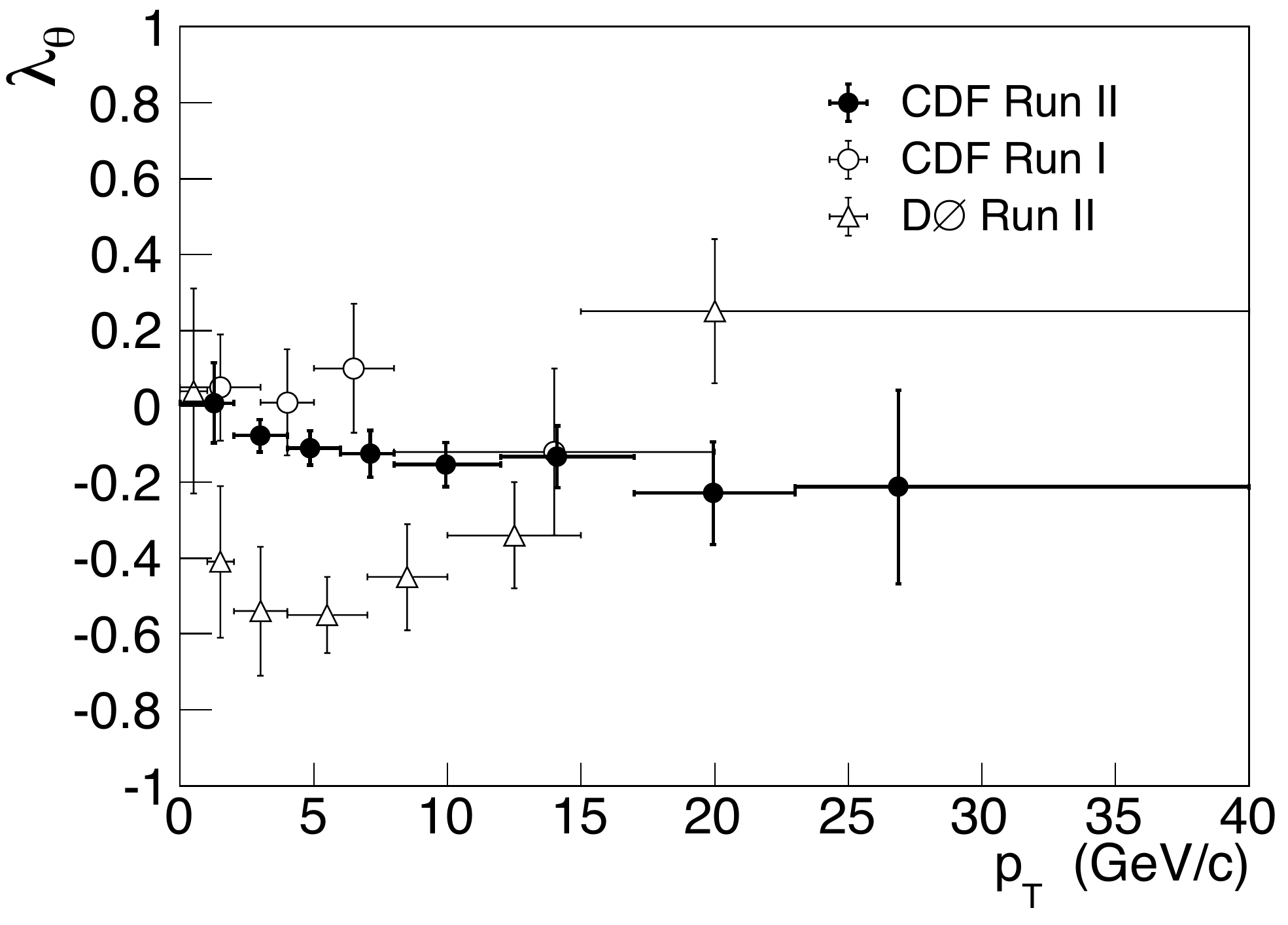}}
\caption{Measurements of $\lambda_{\theta}$ in the cm-helicity frame 
for the $\Upsilon(1S)$ for the three Tevatron experiments: CDF Run~1, CDF Run~2, and D0 Run~2. 
\label{tev_ypol}}
\end{figure}

The Tevatron experiments are summarized in Fig.~\ref{tev_ypol}, 
taken from Ref.~\cite{cdf_run2_ypol}.  For $p_T < 10$~GeV, the CDF 
Run~1 and CDF Run~2 results are statistically consistent, 
while the D0 measurement is radically different.  Note that only the 
CDF Run~2 experiment has employed the 
$\tilde{\lambda}$ systematic uncertainty test to validate its 
results internally.  The consistency of that test, the large 
statistical weight of the sample, and the independent confirmation 
from CDF Run~1 tends to argue that the D0 $\lambda_{\theta}$ results 
are outliers.  The previous discussion of the D0 experiment noted
 that background subtraction was difficult
because of the poor mass resolution and the limited background region
 on the low mass side of the signal region.  

The CDF Run~2 data are the first good-statistics measurements 
of the $\Upsilon(2S)$ and $\Upsilon(3S)$ polarizations.   
None of the three polarization parameters for the higher-mass 
S-wave states shows any significant $p_T$ structure in either 
the cm-helicity or CS frame for $p_T < 10$~GeV.  

\subsubsection{Polarization Results for $\Upsilon(nS)$ with $p_T > 10$~GeV}

\begin{figure}
\centerline{ \includegraphics*[width=16cm,clip=true]{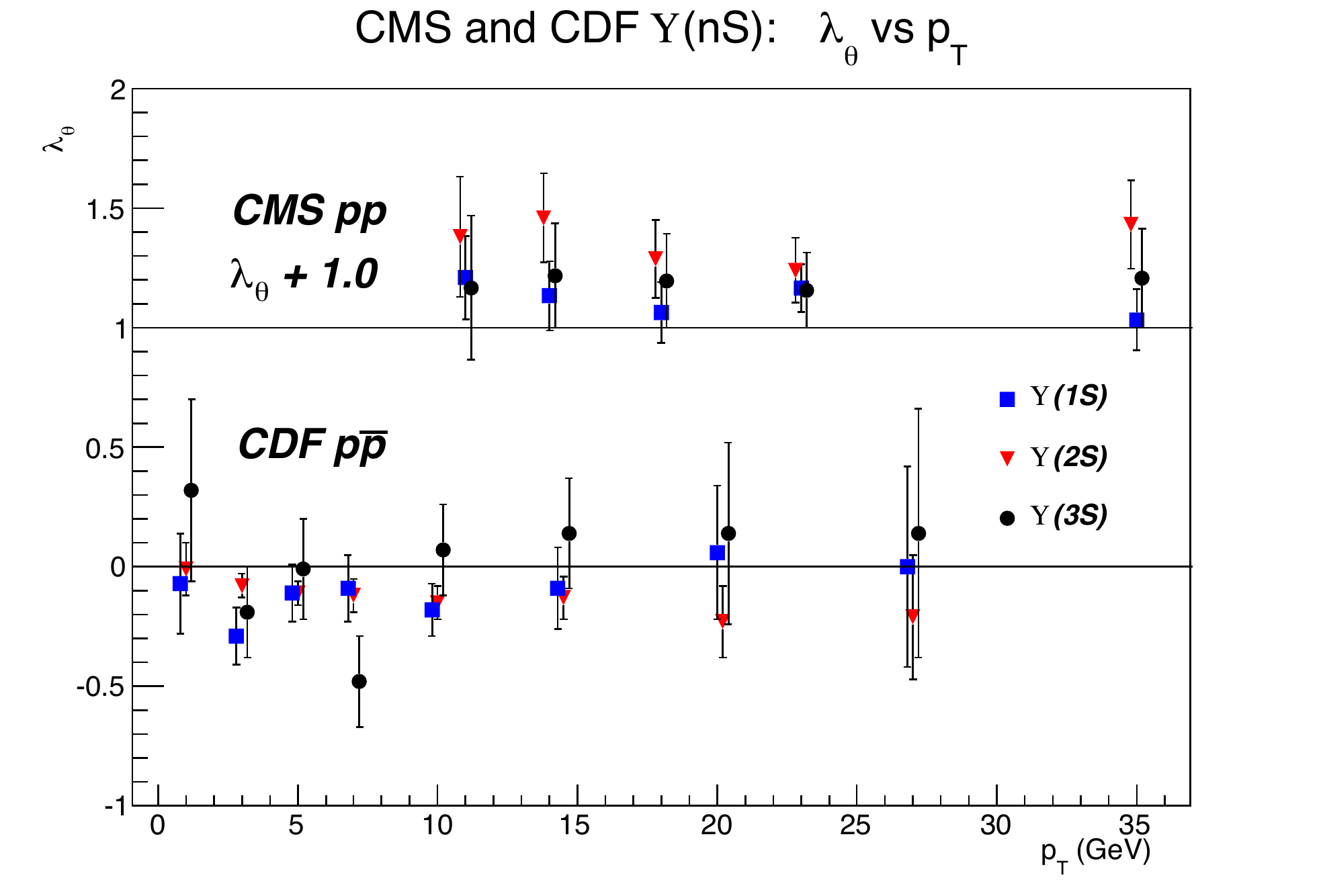}}
\caption{Comparison of $\lambda_{\theta}$ parameters in the 
cm-helicity frame for $\Upsilon(nS)$ production from CDF Run~2 
for $p\overline{p}$ production and from CMS for $pp$ production.
For clarity of presentation, the CMS values have had 1.0 added to 
each $\lambda_{\theta}$ measurement.  Also, for both 
sets of data, the 
$p_T$ values for $\Upsilon(2S)$ and $\Upsilon(3S)$ 
have been shifted left and right by 0.2 GeV, respectively.   
\label{cmscdfcomp}} 
\end{figure}

The $p_T$ range for the CDF Run~2 measurements extends to 40~GeV.  
At the LHC, the CMS collaboration has measured $\Upsilon(nS)$ 
polarizations in the range $10 < p_T < 50$~GeV.  We can compare the 
CMS measurements for $|y| < 0.6$ with the CDF Run~2 results to look 
for possible $p$ versus $\overline{p}$ target effects.  
We compare the $\lambda_{\theta}$ parameter in the cm-helicity frame 
for all three $\Upsilon(nS)$ states in Fig.~\ref{cmscdfcomp}.  
The general features of the two measurements show only 
a small $p_T$ variation of $\lambda_{\theta}$ for $p_T > 10$~GeV. 
The $\Upsilon(1S)$ polarization parameter is relatively more negative
 in both cases, but the statistical uncertainties preclude 
any definite statements about depolarization of the ground state.  
There is a suggestion of an offset between the $p\overline{p}$ 
and $pp$ polarization parameters in the $\Upsilon$ case, 
most clearly in the $\Upsilon(1S)$ case.  This may indicate a 
dependence on having a $p$ or $\overline{p}$ target.

\begin{figure}
\centerline{ \includegraphics*[width=16cm,clip=true]{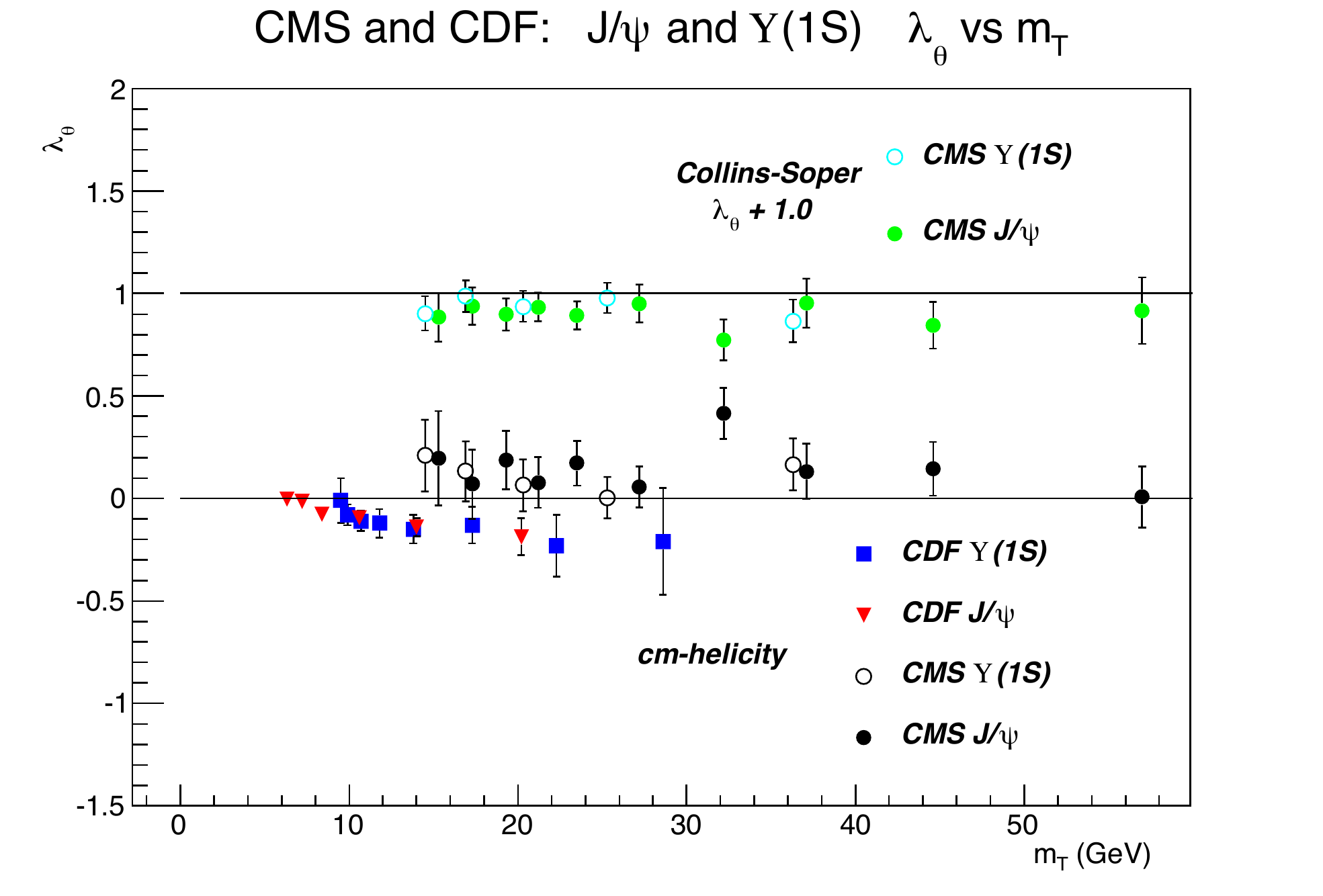} }
\caption{Comparison of $\lambda_{\theta}$ parameters in the
cm-helicity and Collins-Soper frames as functions of the transverse mass $m_T$ 
for $J/\psi$ and for $\Upsilon(1S)$.  The data are from CDF Run~2 and from CMS.  For clarity of presentation, the Collins-Soper values have had 1.0 added to
each $\lambda_{\theta}$ measurement. \label{jycomp}}
\end{figure}

In Fig.~\ref{jycomp}, we plot the $\lambda_{\theta}$ parameters for the
 CMS and CDF measurements in the cm-helicity frame for the $1S$ states 
of bottomonium and charmonium as a function of the transverse mass 
$m_T = \sqrt{m^2 + p_T^2}$.  
We also plot $\lambda_{\theta}$ measurements  
for $J/\psi$ and $\Upsilon(1S)$ in the CS frame as a function of 
$m_T$ from CMS.  The CDF $J/\psi$ data are not available in the CS 
frame.  One sees two features in this figure: 
(a) in each experiment, the polarization parameters of the two onia 
ground states are nearly the same and show the same trend with $m_T$;
 and (b) the trends in the cm-helicity frame are different for $pp$ 
and $p\overline{p}$.  
The $pp$ results are consistent with no $m_T$ dependence 
and a constant $\lambda_{\theta} = 0.13 \pm 0.04$.  
The $p\overline{p}$  $\lambda_{\theta}$ parameter 
shows a linear decrease starting at zero near $m_T$ = 7 GeV with a slope 
of ($-0.015 \pm 0.003$)/GeV.  Again, this {\it may} indicate a 
target-dependent effect that makes the polarization 
different for $pp$ and $p\overline{p}$.  For both types of target, 
the polarization mechanism at large $m_T$ seems to be independent of 
heavy-quark flavor, since the $J/\psi$ and $\Upsilon(1S)$ 
polarization parameters follow the same pattern for each target particle.

\subsection{Discussion}

We have seen some interesting systematic features of quarkonium 
polarization emerge from the comparison of the many available 
measurements at low $p_T$ and from the large-$p_T$ collider 
experiments.  To reiterate, they include
\begin{itemize}
\item 
With the exception of the NuSea pCu bottomonium measurement, 
all of the measured polarization parameters in the cm-helicity or CS frames  
from any experiment are small for the $p_T$ range of 1--70~GeV.
  Furthermore, there is little $p_T$-variation among the measurements
from $pp$ collisions within the uncertainties.
\item 
For $p_T < 10$~GeV, it is striking that the $J/\psi$ polarization parameter 
$\lambda_{\theta}$ in the cm-helicity frame shown in Fig.~\ref{jpollow} is almost 
independent of target particle or rapidity range and tends to be 
$p_T$--independent for $1 < p_T <  10$~GeV.  Those experiments 
that measured $\lambda_{\theta}$ in the CS frame found it to be also 
small, so the polarization in this $p_T$ range is not large in any reference frame.
\item 
As shown in Fig.~\ref{cmscdfcomp} for the cm-helicity frame, the 
$\lambda_{\theta}$ measurements for the 
three $\Upsilon(nS)$ states show little variation with $p_T$ 
or principal quantum number $n$ in either $pp$ or $p\overline{p}$ interactions.
\item 
At comparable $m_T$ values, the polarization parameters for 
ground-state charmonium ($J/\psi$) and ground-state bottomonium 
($\Upsilon(1S)$) are consistent with each other and show little 
variation for $m_T >$ 10 GeV.  We had noted earlier that the 
measured feeddown 
fractions in $p\overline{p}$ experiments for the two quarkonium 
ground-state systems are equal within measurement uncertainties.  
\item 
For $\Upsilon(nS)$ production, there is a suggestion of a difference in
polarization parameters between $pp$ measurements from CMS 
and $p\overline{p}$ measurements from CDF.
\end{itemize}

\section{SUMMARY}

\subsection{Future Experimental Directions}

The question of to what extent feeddown influences the polarization 
of the lowest-lying quarkonium states has been raised repeatedly.  
The best chance to measure these effects seems to be in the large 
datasets collected at the LHC.  Colliding beam experiments at the 
Tevatron and LHC have identified radiative decays of $P$-wave 
quarkonium states to the $S$-wave ground state
using photon conversions in the material of the inner tracker.  
With larger data sets yet unanalyzed,
 one might hope to measure the polarization of the ground-state 
quarkonia that result from $P$-wave decay sources.  We see 
in Fig.~\ref{jycomp} that the polarization parameters of the 
$\Upsilon$(1S) and $J/\psi$ mesons for $m_T >$ 10 GeV are 
consistent with each other and have little variation with $p_T$ in 
either the cm-helicity or CS frames. A first step in understanding
 feeddown effects would be to measure the polarization of the $J/\psi$
 produced from $\chi_c$ decays and of $\Upsilon(1S)$ produced from 
$\chi_{b}(1P)$ decays, 
using conversion photons combined with reconstructed dimuon events 
to identify the $P$-wave parent event candidates.  One need not 
separate the $\chi_{QJ}$ states with different $J$.  
There are already measurements of the $J=2$ to $J=1$ ratios for the $\chi_{c}$ 
at the Tevatron and LHC.  We encourage the 
experimenters to pursue the determination of the $J/\psi$ 
polarization in $\chi_c$ events and to extend the studies to measure the 
$\Upsilon(1S)$ polarization in $\chi_b$ radiative decays.
From the results that we have seen, 
it should be adequate to measure $\lambda_{\theta}$ with respect to two 
spin-quantization axes, the CS frame and either the cm-helicity 
frame or, if the measurements extend out to large rapidity, the $\perp$-helicity frame.
Because one is not measuring a cross section but rather a 
ratio of longitudinal and transverse polarization contributions, the
 absolute photon conversion efficiency is not needed.  A good 
determination of the energy dependence of the conversion efficiency 
is crucial, though, to handle the range of photon energies for 
candidate events.  The fundamental question is whether the 
polarization from these decays is different from the inclusive prompt 
polarization for the $J/\psi$ or $\Upsilon(1S)$.  The answer will 
directly aid future theoretical analysis of quarkonium polarization.

The experimental comparisons of the present data suggest some 
additional studies using existing data.  They include:
\begin{itemize}
\item A new CDF measurement of $J/\psi$ and $\psi(2S)$ polarization 
could gain an order of magnitude more statistics if it were redone 
using the full Tevatron data set.
The increased statistics would allow the determination of all three 
polarization parameters in the analysis, and results could be reported in 
several frames.  It should also be possible to measure the $\psi(2S)$ polarization  
parameters with the larger data set.  It is not clear if D0 has sufficient 
mass resolution to do such a study, but it would be a useful check 
if it were possible.
\item LHCb can lower the measurement uncertainty on its smallest 
rapidity bin using the complete LHC dataset.  This would help to 
evaluate a possible change of polarization with rapidity that cannot
 be excluded by the present measurements.
\item CMS can increase the statistics for its $J/\psi$ polarization 
measurement to decrease the lower $p_T$ cutoff of its measurement, 
especially for a range of rapidity closer to the LHCb lower limit of
 $y = 2$, to investigate the polarization behavior in this potential
 transition region.  Also, reducing the uncertainty on the measurements
 would address the question of whether there is a target dependence in 
the polarization parameters between $pp$ and $p\overline{p}$ 
measurements.
\end{itemize}

The Tevatron and LHC experiments have developed impressive analysis 
techniques and have detectors that work extremely well for the subtle 
business of analyzing polarization.  Applying these tools to available
 data could go far in helping to understand the details of 
polarization in the production of quarkonium in hadronic collisions.

\subsection{Theory Outlook}

The predictions for quarkonium polarization from NRQCD factorization 
at NLO are not in dramatic disagreement with the data,
but the differences are in many cases significant,
given the current experimental and theoretical error bars.
As the experimental uncertainties decrease with higher statistics,
accommodating the data will be increasingly challenging for theory.
As the range of the measurements is extended to higher $p_T$,
there is still an opportunity for theory to predict the polarization.

The NRQCD factorization approach to quarkonium production has
been pushed to NLO in $\alpha_s$, thanks to heroic NLO calculations of the 
pQCD cross sections by three groups independently.
The predictions for polarization at NLO differ dramatically from those at LO.
This raises the question of whether N$^2$LO corrections could be important.
Unfortunately, the calculation of the pQCD cross sections at N$^2$LO
may be prohibitively difficult.

The NLP fragmentation formula, in conjunction 
with the NRQCD expansion of the fragmentation functions,
provides a new framework for quarkonium production at large $p_T$.
Predictions for quarkonium production, 
with pQCD cross sections calculated to NLO in $\alpha_s(p_T)$
and fragmentation functions calculated to NLO in $\alpha_s(m_Q)$,
should be available soon.
It will be interesting to see how the predictions for polarization 
compare to those from NRQCD factorization at NLO.
Since this approach separates the scales $p_T$ and $m_Q$,
reducing calculations of the pQCD cross sections 
and the fragmentation functions to single-scale problems,
calculations to N$^2$LO in $\alpha_s$ may be tractable.

Quantitative predictions of the polarization depend on the choice of data 
used to determine the NRQCD matrix elements.  The safest choices from a
theoretical perspective are data involving the largest $p_T$'s. 
If the data are restricted to spin-summed cross sections at the
large $p_T$'s that are accessible only at the Tevatron and the LHC,
the error bars on polarization predictions are very large.
If polarization measurements are included in the fitting data,
there is still some predictive power in the dependence of the 
polarization on $p_T$.
Testing these predictions requires measurements out to the largest values
of $p_T$ possible.

Current polarization measurements are for the inclusive production of quarkonium.
The sum over all additional hadrons, together with the integration 
over parton momentum fractions, tends to wash out the polarization signal.
The polarization signal could be enhanced by taking into account 
more information about the final-state hadrons, such as the 
direction of the 
hardest jet that balances most of the transverse momentum~\cite{Braaten:2008xg}.  In associated production of quarkonium with another particle, such as a $Z^0$~\cite{Gong:2012ah}, one could also exploit the momentum vector of the associated particle.

\subsection{Concluding Remarks}

The polarization studies from the wide range of experiments covered 
in this review produce a surprisingly coherent picture of quarkonium 
polarization over a wide range of $p_T$.  No experiment observes 
large polarization in any reference frame for either quarkonium flavor 
(except NuSea in pCu collisions).  Nevertheless, the polarization 
parameters in the high-precision experiments (CDF Run~2, CMS) 
are not zero.  The theoretical treatment of polarization is on its 
firmest footing at very large $p_T$.
There are  opportunities at the LHC to extend the 
present measurements into an even higher $p_T$ range, as well 
as to improve the measurement precision by having larger datasets.
In conjunction with theoretical improvements, 
they may allow us to finally develop a clear picture 
of how quarkonium states are produced in hadronic collisions.

\section*{Acknowledgements}

James Russ would like to acknowledge the contributions of his 
colleagues in the CMS and CDF collaborations in many fruitful 
discussions.  His work was supported in part by the U.S.\ Department 
of Energy under Grant No.\ DE-SC0010118TDD. 
Eric Braaten would like to acknowledge discussions with Geoff Bodwin.
Braaten's work was supported in part by the U.S.\ Department 
of Energy under grant DE-FG02-05ER15715.

\end{document}